\documentclass[review]{elsarticle}
\usepackage{mathtools}
\usepackage{float}
\usepackage{multirow}
\usepackage{amsthm}
\usepackage{graphicx}
\usepackage[export]{adjustbox}
\usepackage{changepage}
\usepackage{url}
\usepackage{amssymb}
\usepackage{ulem}
\usepackage{array}
\usepackage{booktabs}
\usepackage[table]{xcolor}
\usepackage{hyperref}

\hypersetup
{
  colorlinks   = true, %Colours links instead of ugly boxes
  urlcolor     = red,  %Colour for external hyperlinks
  linkcolor    = blue, %Colour of internal links
  citecolor    = blue  %Colour of citations
}
\usepackage{wasysym}
\usepackage{listings}
\usepackage{pdflscape}
\usepackage{rotating}
\usepackage[utf8]{inputenc}
\usepackage[T1]{fontenc}
\usepackage{tikz,lipsum,lmodern}
\usepackage[most]{tcolorbox}
\usepackage{pgfplots}
\usepackage[toc,page]{appendix}
\usepackage{textcomp}
\usepackage{epstopdf}
\usepackage{csquotes}
\usepackage{dirtytalk}
\usepackage{parskip}
\usepackage{amssymb}
\usepackage{tikz}
\usepackage{comment}
\usepackage{indentfirst}% this helps to indent the first paragraph after the heading or chapter
\usepackage[colorinlistoftodos,prependcaption,textsize=tiny]{todonotes}
\usepackage[english]{babel}
\usepackage{geometry}
\pgfplotsset{compat=1.9}
\usepackage{lineno}
\usepackage{caption} 
\usepackage{adjustbox}
\usepackage{pdfpages}
 \usepackage{ragged2e} 
\usepackage{xurl}
\usepackage{longtable,tabu}
\usepackage{afterpage}
\usepackage{verbatim}
\usepackage{subfigure}
\usepackage{subfloat}
\usepackage{longtable}
\usepackage{ltxtable}
\usepackage{fancyvrb}
\usepackage{tabularx}
\usepackage{makecell}
\usepackage{supertabular}

\journal{Elsevier}

\begin{document}

\begin{frontmatter}

\title{An exploratory study on automatic identification of assumptions in the development of deep learning frameworks}




\author[mythirdaddress,myfourthaddress]{Chen Yang}
\ead{yangchen@szpu.edu.cn}
\address[mythirdaddress]{School of Artificial Intelligence, Shenzhen Polytechnic University, 518055 Shenzhen, China}
\address[myfourthaddress]{State Key Laboratory for Novel Software Technology, Nanjing University, Nanjing, China}

\author[mymainaddress]{Peng Liang\corref{mycorrespondingauthor}}
\cortext[mycorrespondingauthor]{Corresponding author at: School of Computer Science, Wuhan University, China. Tel.: +86 27 68776137; fax: +86 27 68776027.}
\ead{liangp@whu.edu.cn}
\address[mymainaddress]{School of Computer Science, Wuhan University, 430072 Wuhan, China}

\author[mysecondaddress]{Zinan Ma}
\ead{zinanm@chalmers.se}
\address[mysecondaddress]{Department of Computer Science and Engineering, Chalmers University of Technology, S-412 96 Gothenburg, Sweden}

\begin{abstract}
\justifying
\textbf{Context}: Stakeholders constantly make assumptions in the development of deep learning (DL) frameworks. These assumptions are related to various types of software artifacts (e.g., requirements, design decisions, and technical debt) and can turn out to be invalid, leading to system failures. Existing approaches and tools for assumption management usually depend on manual identification of assumptions. However, assumptions are scattered in various sources (e.g., code comments, commits, pull requests, and issues) of DL framework development, and manually identifying assumptions has high costs (e.g., time and resources).\\
\textbf{Objective}: The objective of the study is to evaluate different classification models for the purpose of identification with respect to assumptions from the point of view of developers and users in the context of DL framework projects (i.e., issues, pull requests, and commits) on GitHub. \\
\textbf{Method}: First, we constructed a new and largest dataset (i.e., the AssuEval dataset) of assumptions collected from the TensorFlow and Keras repositories on GitHub. Then we explored the performance of seven non-transformers based models (e.g., Support Vector Machine, Classification and Regression Trees), the ALBERT model, and three decoder-only models (i.e., ChatGPT, Claude, and Gemini) for identifying assumptions on the AssuEval dataset.\\
\textbf{Results}: The study results show that ALBERT achieves the best performance (f1-score: 0.9584) for identifying assumptions on the AssuEval dataset, which is much better than the other models (the 2nd best f1-score is 0.8858, achieved by the Claude 3.5 Sonnet model). Though ChatGPT, Claude, and Gemini are popular models, we do not recommend using them to identify assumptions in DL framework development because of their low performance. Fine-tuning ChatGPT, Claude, Gemini, or other language models (e.g., Llama3, Falcon, and BLOOM) specifically for assumptions might improve their performance for assumption identification.\\
\textbf{Conclusions}: This study provides researchers with the largest dataset of assumptions for further research (e.g., assumption classification, evaluation, and reasoning) and helps researchers and practitioners better understand assumptions and how to manage them in their projects (e.g., selection of classification models for identifying assumptions).
\end{abstract}


\end{frontmatter}

\section{Introduction} \label{introduction} 
Deep learning (DL) is a branch of machine learning (ML), which has been widely used in various areas as a key component, such as image classification, question-and-answer, and speech recognition. To improve the efficiency of developing DL-based applications, many popular DL frameworks (e.g., Theano~\cite{theano2022}, Caffe~\cite{jia2014caffe}, TensorFlow~\cite{Abadi2016}, and PyTorch~\cite{Steiner2019}) have been designed, implemented, and used in DL systems. Due to lack of knowledge or information, time pressure, complex context, etc., various uncertainties emerge during the development of DL frameworks, leading to assumptions made in projects \cite{Yangsms2018}. 

In this work, we adopt the definition of the assumption concept from our previous work \cite{Yangsms2018}: ``\textit{software assumptions are software development knowledge taken for granted or accepted as true without evidence.}'' Such a definition of assumptions emphasizes the characteristic of uncertainty in software development knowledge: stakeholders believe but cannot know for sure the importance, impact, correctness, etc. of specific software development knowledge (e.g., requirements knowledge and design knowledge). 
Moreover, as defined by Kroll and Kruchten \cite{Kroll2003}: ``\textit{an artifact is a piece of information that is produced, modified, or used by a process}''. According to this definition, assumptions are a type of software artifacts, similar to requirements, design decisions, etc.

The assumptions in DL frameworks are different from those of traditional software systems. In addition to assumptions about system requirements, design, and components, many assumptions in DL frameworks are about data, models, and training~\cite{siebert2021}. For example, in the documents (e.g., \texttt{multigpu.md}) of Caffe, developers explicitly mention assumptions, such as assuming that GPUs for training are of the same type\footnote{\url{https://github.com/BVLC/caffe/blob/master/docs/multigpu.md}}. Another example is that Erickson \textit{et al.} pointed out that the assumptions of the input data dimensions in TensorFlow and Theano are different, and this should be considered in system design and coding when using such DL frameworks~\cite{Erickson2017}.

\begin{figure}[!h]
\centering    
\includegraphics[scale=0.2]{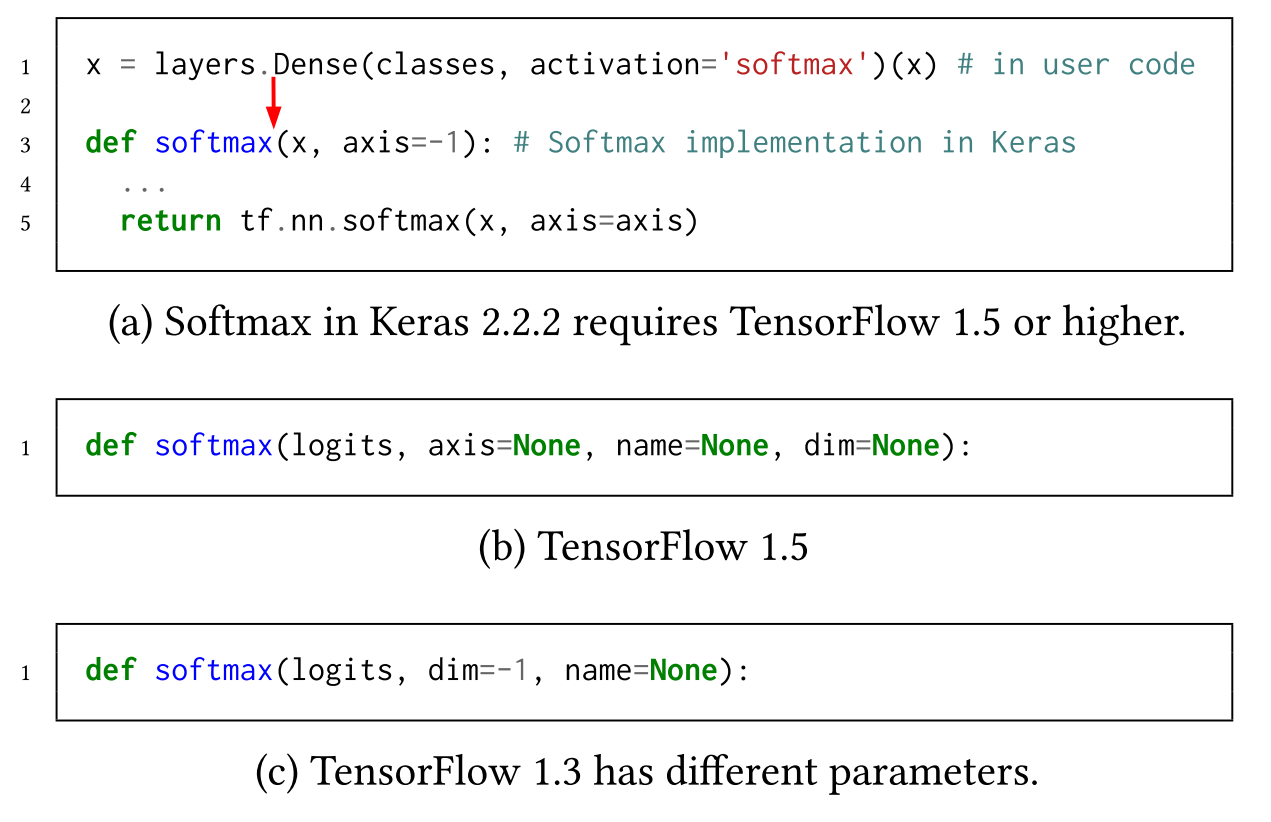}
\caption{An invalid assumption of Softmax implementation in Keras 2.2.2 and call in TensorFlow 1.3 and 1.5~\cite{Zhang2020}}
\label{fig_tf_keras_evolution}
\end{figure}

The importance of assumptions and their management in software development (including DL framework development) has been highlighted in many studies. For example, Corbató \cite{Corbato1991} mentioned in his ACM Turing Award lecture that ``\textit{design bugs are often subtle and occur by evolution with early assumptions being forgotten as new features or uses are added to systems.}'' Garlan \textit{et al.} pointed out that incompatible assumptions in software architecture can cause architectural mismatch \cite{Garlan2009}. Lewis \textit{et al.} also got similar results in ML systems: since there are different types of stakeholders (e.g., data scientist, software engineer, and system user) of an ML system, they could make different but incompatible or invalid assumptions, leading to system misunderstanding, mismatch, etc. \cite{lewis2021}. 
Zhang \textit{et al.}~\cite{Zhang2020} collected 4,960 real failures of several DL frameworks, such as TensorFlow, PyTorch, and Keras, and found that part of system failures are caused by invalid assumptions. 
For example, during the evolution of TensorFlow and Keras, stakeholders assumed that Keras 2.2.2 can be used with TensorFlow 1.3 without being aware of the changes in the Softmax function in Keras 2.2.2, as shown in Figure \ref{fig_tf_keras_evolution}, which led to system failures.

Many industrial cases also indicate the importance of assumptions and their management. 
For example, in two accidents involving the Boeing 737 MAX aircraft, significant tragedies occurred. In October 2018, Lion Air Flight 610 crashed 13 minutes after take-off and killed all 189 people on board; in March 2019, Ethiopian Airlines Flight 302 crashed and ended another 157 lives.
According to government reports, a critical reason for Boeing 737 MAX crashes is related to poorly managed assumptions \cite{US2019}\cite{committee2020}. The reports indicated that the aircraft manufacturer made invalid assumptions about critical system components. Specifically, invalid assumptions regarding the MCAS (Maneuvering Characteristics Augmentation System) are the root cause of crashes. The reports also highlighted the need to identify and re-evaluate important assumptions in the system. 

Assumptions are related to many types of software artifacts, such as decisions, technical debt, and source code \cite{Yangsca2021}. For example, in TensorFlow, there is an assumption: ``\textit{TODO: Looks like there is an assumption that weight has only one user. We should add a check here}'', which induces a technical debt issue: a check should be added on whether the weight has a single user in the future.
Existing research on assumptions and their management in software development usually uses experiments, surveys, and case studies to manually identify and extract assumptions through observation, questionnaires, interviews, focus groups, and documentation analysis~\cite{Yangsms2018}. Since such approaches are highly cost-effective, for example, in terms of time and resources, the comprehensiveness of the identified and extracted assumptions in these studies is often limited, leading to various problems in developing new theories, approaches, and methods of assumptions and their management in software development.
In this work, to overcome the issue of manually identifying assumptions in DL framework development, we conducted an exploratory study to find a good algorithm and model to identify assumptions on GitHub. This provides a foundation for further research on assumptions and their management in the development of DL frameworks, for example, assumption classification, documentation, evaluation, and reasoning ~\cite{Yangsms2018}.

The \textbf{main contributions} of this study are as follows.

(1) We designed a set of applicable criteria to identify assumptions and constructed a new and largest dataset (i.e., AssuEval \cite{yangpackage2023}) of assumptions collected from GitHub projects. 

(2) We evaluated the performance of seven non-transformers based models, the ALBERT model, and three decoder-only models (ChatGPT, Claude, and Gemini) to identify assumptions on the AssuEval dataset.

The remainder of the paper is organized as follows. 
Section \ref{Background} and Section \ref{relatedwork} provide background knowledge and related work, respectively, Section \ref{assuevaldataset} clarifies the AssuEval dataset, Section \ref{design} presents the evaluation design for different classification models, Section \ref{results} presents the results, Section \ref{discussion} discusses the results, Section \ref{threats} clarifies the threats to the validity of the study, and Section \ref{conclusions} concludes the work with future directions.

\section{Background} \label{Background}
\subsection{Assumptions in software development}
During software development, there can be many uncertain things. However, in order to meet the project business goals (e.g., schedule and deadlines), stakeholders have to work in the presence of such uncertainties; these uncertainties can lead to assumptions. For example, uncertainty regarding the release date of a specific technology to be used in a system may lead to making an assumption about that release date. In this study, we advocate treating uncertainty and assumption as two different but related concepts: one way to deal with uncertainties is to make implicit or explicit assumptions, but not all uncertainties lead to assumptions. In addition to the uncertain nature of assumptions, we summarize the other four main characteristics of assumptions in software development:

\textbf{Subjective}: Many researchers and practitioners pointed out that whether a piece of information is an assumption or not is rather subjective (e.g., \cite{Yangsms2018}\cite{Yang2017}\cite{Roeller2006}\cite{Wang2016}). This is the major reason that stakeholders may have a different understanding of the assumption concept. Many studies also mention that it is difficult to draw a line between assumptions and other types of software artifacts. As an example, Roeller \textit{et al.}~\cite{Roeller2006} mentioned: ``\textit{from one perspective or stakeholder, we may denote something as an assumption, while that same thing may be seen as a design decision from another perspective.}''

\textbf{Intertwined with certain types of artifacts}: Assumptions are not independent in software development, but intertwined with many other types of software artifacts. For example, when managing assumptions in software design (e.g., \cite{Yangsms2018}\cite{Landuyt2012}\cite{Tang2018}), assumptions are commonly related to requirements, design decisions, components, etc.

\textbf{Dynamic}: Assumptions have a dynamic nature, i.e., they can evolve over time \cite{Yang2017}\cite{Lewis2004}. For example, during software development, a valid assumption can turn out to be invalid or vice versa, or an assumption can transform to another type of software artifact or vice versa.

\textbf{Context-dependent}: Assumptions are context-dependent \cite{Yang2017}. For example, the same assumption could be valid in one project and invalid in another project because the context changes; or an assumption in one project is not an assumption in another project. Unless the information is expressed in an explicit way (e.g., using phrases such as ``\textit{it is assumed that''}), it is difficult to judge whether the information is an assumption or not, without considering its context \cite{Yang2017}.

\subsection{Large language models}
In this section, we introduce two types of popular large language models, i.e., encoder-only models including BERT and ALBERT and decoder-only models including GPT, Claude, and Gemini. 
The architecture of the BERT model is based on the original Transformer model \cite{Vaswani2017}, which is a multilayer bidirectional Transformer encoder \cite{Devlin2019}. The BERT model was designed for multitasking, such as natural language understanding, question and answer, and sentence completion. In the original paper, Devlin \textit{et al.} proposed two versions of the BERT model: BERT-Base and BERT-Large \cite{Devlin2019}. BERT-Base has 12 transformer blocks and 12 self-attention heads, the hidden size is 768, and the total parameters are 110M; BERT-Large has 24 transformer blocks and 16 self-attention heads, the hidden size is 1024, and the total parameters are 340M \cite{Devlin2019}. To better deal with environments with restricted computational resources, Turc \textit{et al.} proposed different versions of the BERT model with smaller sizes, for example, BERT-Tiny, BERT-Mini, BERT-Small, and BERT-Medium, which reduce the number of transformer blocks and hidden size \cite{turc2019}.

\begin{figure}[htbp]     
\centering    
\includegraphics[scale=0.17]{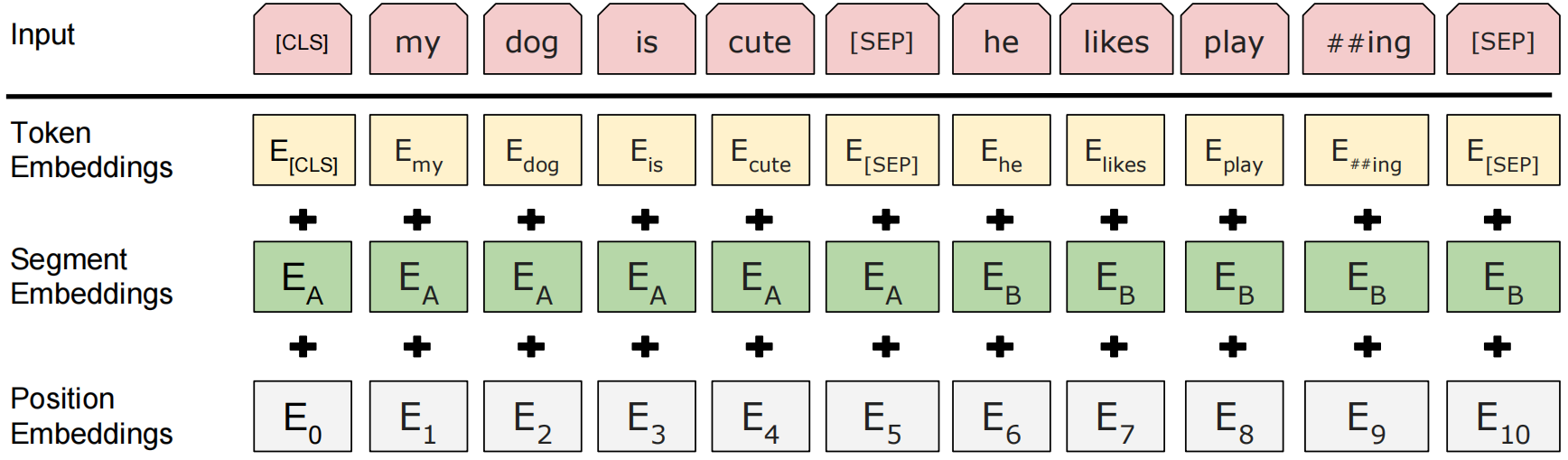}
\caption{BERT input embeddings \cite{Devlin2019}}
\label{bert_embedding}
\end{figure}

The input of the BERT model accepts a sentence or pair of sentences in a token sequence to allow the model to perform different tasks \cite{Devlin2019}. Devlin \textit{et al.} used WordPiece embeddings with a vocabulary of 30,000 tokens \cite{Devlin2019}. In the embeddings, they used a specific token: \textit{[SEP]} to separate the two sentences and added a learned embedding to each token to show whether a token belongs to the first sentence or the second \cite{Devlin2019}. Taking into account an input that includes two sentences: ``\textit{my dog is cute}'' and ``\textit{he likes playing}'', the input embeddings are the sum of the token embeddings, the segmentation embeddings, and the position embeddings, as shown in Figure \ref{bert_embedding}.

Since the trained BERT model can successfully deal with a broad set of NLP (Natural Language Processing) tasks, many researchers tried to adapt the original BERT model, leading various models, such as XLNet \cite{YangXlnet2019}, RoBERTa \cite{LiuRoBERTa2019}, ALBERT \cite{LanALBERT2020}, and ELECTRA \cite{ClarkELECTRA2020}. These models focus on different aspects of improving the original BERT model. In this paper, we chose one of the most representative models: ALBERT, which aims to reduce parameters, lower memory consumption, and increase the training speed of the BERT model \cite{LanALBERT2020}.
Compared to the BERT model, ALBERT has a similar architecture but uses factorized embedding parameterization, cross-layer parameter sharing, and inter-sentence coherence loss for optimization \cite{LanALBERT2020}. 
The ALBERT model has four versions: ALBERT-Base (12M parameters), ALBERT-Large (18M parameters), ALBERT-xLarge (60M parameters), and ALBERT-xxLarge (235M parameters), with much fewer parameters than the BERT model, but did better in certain tasks (e.g., GLUE, RACE, and SQuAD benchmarks) \cite{LanALBERT2020}.

The GPT models are based on the decoder-only transformer architecture \cite{Minaee2024} and exhibit a distinctive methodology for language processing: rather than engaging in the interpretation or analysis of text in which Encoder-Only Models excel, Decoder-Only models are primarily dedicated to the generation of new text.
The GPT-3.5 series of models includes code-davinci-002, text-davinci-002, text-davinci-003, and gpt-3.5-turbo-0301, which can be traced back to 2020\footnote{\url{https://platform.openai.com/docs/model-index-for-researchers}}. Brown \textit{et al.} discovered that scaling language models could significantly enhance task-agnostic, few-shot performance, sometimes even surpassing the performance of prior state-of-the-art fine-tuning approaches \cite{Brown2020}. Then they proposed GPT-3, which comes in three versions: GPT-3 175B (175 billion parameters), GPT-3 6.7B (6.7 billion parameters), and GPT-3 1B (1 billion parameters). They have demonstrated strong performance on a variety of NLP tasks, including translation, question-answering, cloze tasks, unscrambling words, using a novel word in a sentence, and performing 3-digit arithmetic.
Chen \textit{et al.} proposed Codex: a GPT language model fine-tuned on publicly available code from GitHub \cite{Chen2021}. This model has only one version: Codex 12B (12 billion parameters). To evaluate its Python code-writing capabilities, they created a new dataset called HumanEval, which can be used to measure functional correctness for synthesizing programs from docstrings. In HumanEval, Codex solved 28.8\% of the problems, while GPT-3 solved 0\% of the problems. The authors also discovered that repeated sampling from the model is an effective way to generate working solutions to difficult prompts. With 100 samples per problem, they were able to solve 70.2\% of their problems using the repeated sampling strategy.
Neelakantan \textit{et al.} investigated pretraining techniques \cite{Neelakantan2022}. They discovered that using contrastive pretraining on unsupervised data with a large batch size can generate high-quality vector representations of text and code. Their models achieved the best results on linear-probe classification, text search, and code search, but underperformed on sentence similarity tasks. In their work, they trained the GPT-3 unsupervised cpt-text model, which has three versions: GPT-3 unsupervised cpt-text 175B (175 billion parameters), GPT-3 unsupervised cpt-text 6B (6 billion parameters), and GPT-3 unsupervised cpt-text 1.2B (1.2 billion parameters).
Stiennon \textit{et al.} created a dataset of human-generated comparisons between summaries and then used it to train a model to predict the human-preferred summary \cite{Stiennon2022}. This model was also used as a reward function to fine-tune a summarization policy through reinforcement learning. This model has three versions: GPT-3 6.7B pre-train (6.7 billion parameters), GPT-3 2.7B pre-train (2.7 billion parameters), and GPT-3 1.3B pre-train (1.3 billion parameters). Their study shows that it is possible to significantly improve summary quality by training a model to optimize for human preferences.
Ouyang \textit{et al.} developed a model called InstructGPT, which has three versions: InstructGPT-3 175B (175 billion parameters), InstructGPT-3 6B (6 billion parameters), and InstructGPT-3 1.3B (1.3 billion parameters) \cite{Ouyang2022}. Even the InstructGPT-3 1.3B version, which has much fewer parameters than GPT-3 (175B parameters), outperformed GPT-3 in certain areas (e.g., truthfulness and reductions in toxic output generation). Furthermore, they found that fine-tuning with human feedback is a promising approach to align language models with human intent.
OpenAI published a report of GPT-4 in 2023 \cite{OpenAI2023}. GPT-4 is also based on Transformer, which is the same as the GPT-3 and GPT-3.5 series of models. GPT-4 is capable of taking not only written texts, but also images. The experiments in the report demonstrate that GPT-4 is superior to existing decoder-only models (e.g., GPT-3 and GPT-3.5 models) on a set of NLP tasks and surpasses the majority of the reported state-of-the-art systems.

Similarly to the GPT models, there are other popular decoder-only models such as Claude and Gemini \cite{Minaee2024}\cite{Chang2024}. Claude is a family of decoder-only models developed by Anthropic\footnote{\url{https://www.anthropic.com}}, supporting advanced reasoning, vision analysis, code generation, and multilingual processing. The first Claude model was released in March 2023\footnote{\url{https://www.anthropic.com/news/introducing-claude}}. The Claude models have three branches: Haiku, Sonnet, and Opus. In the Claude family, the Haiku series models are lightweight and fastest models, which can execute lightweight actions with industry-leading speed; the Sonnet series models are hard-working models, which are the best combination of performance and speed for efficient, high-throughput tasks; the Opus series models are the most powerful models, which can process complex tasks. Gemini also has capabilities across text, image, video, and audio \cite{Gemini2024}. The Genimi series models also include three types: Ultra, Pro, and Nano, which is similar to the Claude family \cite{Gemini2024}. Gemini Ultra is the most capable model, which achieved state-of-the-art scores on various tasks; Gemini Pro is balanced between cost and performance; Gemini Nano is the most lightweight model in the Gemini family, which was designed specifically for devices \cite{Gemini2024}.
As mentioned in related studies, the GPT models, Claude models, Gemini models, etc. have evolved over time and can be used for multiple tasks, including identifying assumptions in the development of DL frameworks.

\section{Related work} \label{relatedwork}
In the field of assumptions and their management in software development, most assumptions are manually identified by researchers and practitioners. 
Landuyt and Joosen focused on the assumptions made during the application of a threat modeling framework (that is, LINDDUN), which allows the identification of privacy-related design flaws in the architecting phase \cite{Landuyt2020}. They conducted a descriptive study with 122 master students, and the students identified and extracted 845 assumptions from the models created by the students. 
In our previous work, we conducted an exploratory study as a first step in understanding assumptions in terms of their distribution, classification, and impacts based on code comments from nine popular DL framework projects on GitHub~\cite{Yangsca2021}. The results show that (1) 3,084 assumptions are scattered across 1,775 files in the nine DL frameworks, ranging from 1,460 (TensorFlow) to 8 (Keras) assumptions. (2) There are four types of assumption validity: Valid, Invalid, Conditional, and Unknown, and four types of assumptions based on their content: Configuration and Context, Design, Tensor and Variable, and Miscellaneous. (3) Both valid and invalid assumptions can have an impact within a specific scope (e.g., in a function) on the DL frameworks. Certain technical debt is induced when making assumptions, and source code is written and decisions are made based on assumptions. 
Xiong \textit{et al.} studied the assumptions in the Hibernate Developer mailing list, including their expression, classification, trend over time, and related artifacts \cite{Xiong2018}. In their study, they identified and extracted 832 assumptions. Li \textit{et al.} developed an ML approach \cite{Li2019} to identify and classify assumptions based on the dataset constructed by Xiong \textit{et al.} \cite{Xiong2018}, which can read the data (i.e., sentences) from the dataset (i.e., a \texttt{.csv} file), preprocess the data (e.g., using NLTK and Word2Vec), train specific classifiers (e.g., Perception, Logistic Regression, and Support Vector Machine), and evaluate the trained classifiers (i.e., precision, recall, and F1 score). 
We listed the differences of our work and the related work above in Table \ref{Comparison}.

\begin{table}[!h]
\scriptsize
\centering
\caption{Comparison of our work and the related work}
\label{Comparison}
\begin{tabular}{p{0.13\columnwidth}p{0.15\columnwidth}p{0.1\columnwidth}p{0.1\columnwidth}p{0.1\columnwidth}p{0.25\columnwidth}}
\toprule
\textbf{Study} & \textbf{Dataset} & \textbf{Scale} & \textbf{Identification} & \textbf{Classification} & \textbf{Models}\\
\midrule
Landuyt and Joosen \cite{Landuyt2020} & Student projects & 845 & Manual & Multiclass & NA \\
Yang \textit{et al.} \cite{Yangsca2021} & DL frameworks & 3,084 & Manual & Multiclass & NA \\
Xiong \textit{et al.} \cite{Xiong2018} & Hibernate project & 832 & Manual & Multiclass & NA \\
Li \textit{et al.} \cite{Li2019} & Hibernate project & 832 & Automatic & Binary & Non-transformers based models \\
\textbf{Our work} & DL frameworks & 15,354 & Automatic & Multiclass & Non-transformers based models, the ALBERT model, and Decoder-only models \\
\bottomrule
\end{tabular}
\end{table}

\begin{figure}[h]
    \centering
    \subfigure[Implementation of the Fibonacci sequence calculation]{
        \centering
        \includegraphics[width=0.4\linewidth]{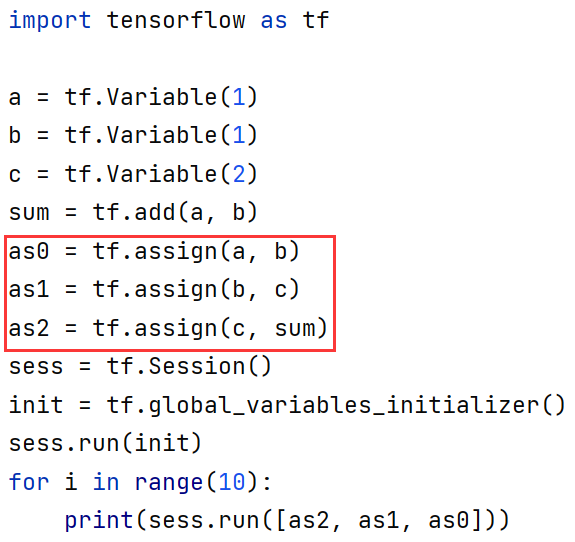}
        \label{fig1}}~~\hfill~
    \subfigure[Arbitrary results of the Fibonacci sequence calculation]{
        \centering
        \includegraphics[width=0.6\linewidth]{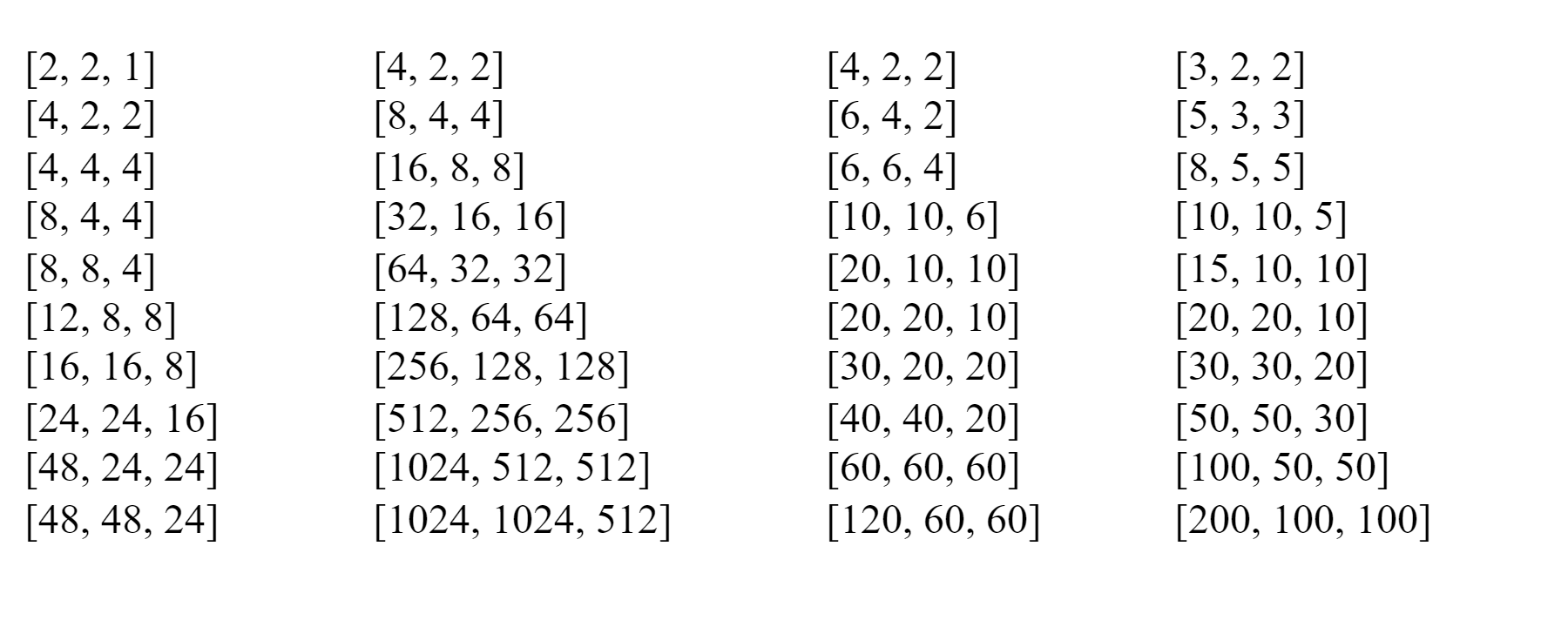}
        \label{fig2}}~~\hfill~
    \caption{An example of TensorFlow misuse caused by an invalid assumption~\cite{Zhang2018}}
    \label{fig_tf_compute_sequence}
\end{figure}

There are also studies that focus on other types of artifacts in the development of DL frameworks and identify assumptions related to the artifacts.
Zhang \textit{et al.}~\cite{Zhang2020} collected 4,960 real failures from a Microsoft DL platform. This platform includes multiple DL frameworks, such as TensorFlow, PyTorch, and Keras. The authors manually examined the failure messages from the 4,960 system failures and classified them into 20 categories. They also identified root causes and solutions in a sample of 400 failures, including system failures caused by assumptions.
Islam \textit{et al.}~\cite{Islam2019} focused on bugs in DL frameworks collected from both Stack Overflow and GitHub, and studied Caffe, Keras, TensorFlow, Theano, and Torch to understand the types, root causes, and impacts of bugs, the bug-prone stage of the DL pipeline, and related antipatterns in the frameworks. They mentioned that some bugs are caused by invalid assumptions of the behavior and specifications for different APIs in different frameworks/libraries. For example, different frameworks may have similar functions, such as adjusting the size of an image (e.g., \textit{PIL} in Keras and \textit{tensorflow.image.resize\_images} in TensorFlow). Users usually need to adjust the image size in multiple components of a DL system and may use different APIs provided by different frameworks interchangeably, assuming that they have the same behavior. However, when such an assumption is invalid, it can cause system failures.
Zhang \textit{et al.}~\cite{Zhang2018} conducted an empirical study on 175 TensorFlow bugs collected from Stack Overflow and GitHub. They studied the root causes and symptoms of bugs and solutions for bug detection and localization. The authors also pointed out that certain system failures are caused by invalid assumptions. For example, they found that stakeholders made an invalid assumption regarding computational models, which caused the construction of an invalid TensorFlow computational graph. Specifically, a TensorFlow user used \textit{tf.assign} to calculate a Fibonacci sequence and defined three nodes as \textit{as0}, \textit{as1}, and \textit{as2}, as shown in Figure \ref{fig1}. The user assumed that the nodes followed the conventional control flow semantics and would be computed sequentially, which is an invalid assumption, since TensorFlow does not impose a sequential order on the computation of the nodes, i.e., the nodes can be computed in an arbitrary order, as shown in Figure \ref{fig2}.

As described in the related studies mentioned above, assumptions and their management are important in the development of DL frameworks. However, existing approaches for assumption identification are typically either time-consuming or have been studied on a limited scale of data.

\section{The AssuEval dataset} \label{assuevaldataset}
In this study, we considered two types of assumptions: self-claimed assumptions (SCAs) and potential assumptions (PAs). The rationale is that a sentence cannot be identified as an assumption unless it is explicitly mentioned by stakeholders, which are SCAs. There are also some plausible cases that may or may not be assumptions, which we have to ask stakeholders to confirm, which are PAs.
Taking into account a sentence in a commit, pull request (PR), or issue of a GitHub repository, if the sentence includes an assumption and the assumption is explicitly claimed using at least one of the assumption-related terms (i.e., ``assumption'', ``assumptions'', ``assume'', ``assumes'', ``assumed'', ``assuming'', ``assumable'', and ``assumably''), we considered that the sentence includes an SCA.
For example, in this sentence ``\textit{[tf/xla] fixup numbering of xla parameters used for aliasing previously, the xla argument parameter was incorrectly assumed to correspond to the index in the vector of `xlacompiler::argument'}'', it includes an SCA of assuming that the \textit{xla argument} parameter corresponds to the index in the vector of \textit{xlacompiler::argument} and the developer claimed that it was an invalid assumption. 
If a sentence may include an assumption, which needs further confirmation from human experts, we considered the sentence includes a PA.
This definition covers various aspects, such as expectations, future events, possibilities, guesses, opinions, feelings, and suspicions, which can indicate PAs.
We provide three examples to further explain the concept of PA. For example, in the sentence ``\textit{I think the right way to create demo tensorboard instances is to simply run a tensorboard in the cloud, rather than keep maintaining this mocked-out backend.}'', it includes a PA regarding thoughts on the right way to create demo tensorboard instances. In another example ``\textit{The system will not crash under heavy load}'', the sentence describes the future state of the system, which is uncertain and includes a PA. The third example ``\textit{both false and true outputs should be considered independently}'' does not include assumption-related terms, but the sentence describes an expectation (i.e., ``something should be''), which is a PA. After further confirmation by human experts, this PA can be transformed into an SCA or other types of software artifacts. 
Besides SCAs and PAs (which belong to explicit assumptions), there are also many implicit assumptions in projects (e.g., in stakeholders' heads or requiring reasoning). Identifying implicit assumptions is much more tricky than identifying SCAs and PAs, since there are no explicit clues in the sentences and stakeholders need to infer the sentences based on the context, which involves assumption reasoning. Identifying implicit assumptions is beyond the scope of this study and we treat it as future work.
Therefore, for a sentence that is neither an SCA nor a PA, we considered it as ``not an assumption'' (NA).

\begin{figure*} [!h]
 \centering
  \includegraphics [scale=0.12] {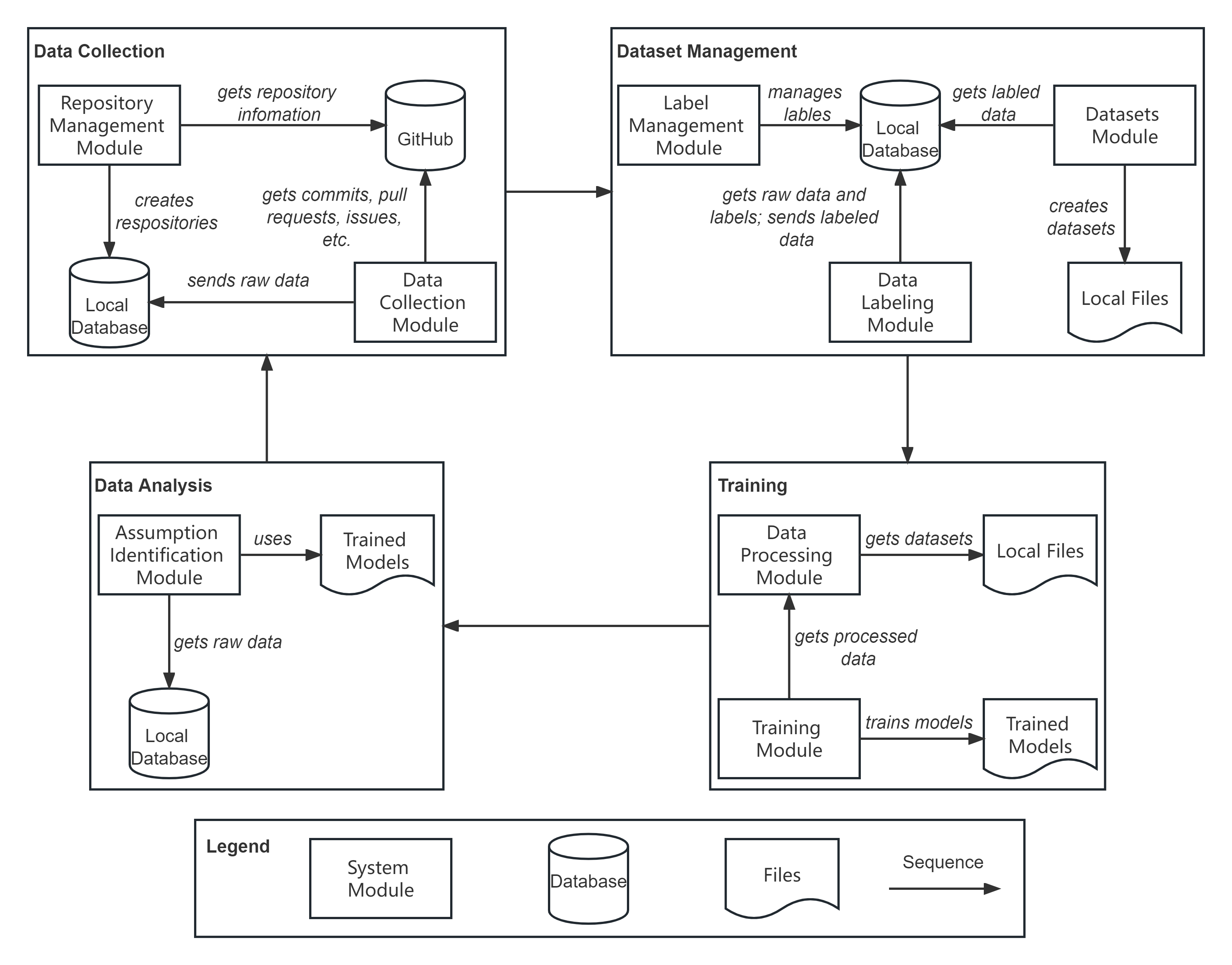}
 \caption{Assumption identification framework for GitHub}
 \label{framework}
\end{figure*}

Before further exploring the performance of the classification models for assumption identification, we need to first create a dataset of SCAs, PAs, and NAs (i.e., the AssuEval dataset). We used a framework and developed a tool named Assumption Miner~\cite{Yangtool2023} that implements the framework to conduct the study. The framework has four components: Data Collection, Dataset Management, Training, and Data Analysis, as shown in Figure~\ref{framework}.
\textbf{Data Collection} aims to (1) get information on repositories and their releases from the GitHub server and (2) collect commits, PRs, issues, etc. from specific repositories. 
\textbf{Dataset Management} includes (1) label management (e.g., adding or editing a label), (2) labeling of raw data (e.g., commits) from specific repositories, and (3) generation of datasets into files (e.g., \texttt{csv} files).
\textbf{Training} first reads and processes data from the generated datasets, constructs the selected models, and then trains the models.
\textbf{Data Analysis} uses the trained models to identify assumptions on new data.
Details of the framework and the tool can be found in the replication package of this study \cite{yangpackage2023}.

\subsection{Data collection}
% The process of data collection is shown in Fig. \ref{data_collection}.
The process of data collection is composed of five steps as described below.

(1) \textbf{Select repositories}. TensorFlow is one of the most popular DL frameworks, which is widely used in many DL systems and application domains. 
The TensorFlow project on GitHub started in 2015, having 161,680 commits\footnote{\url{https://github.com/tensorflow/tensorflow} (accessed on 2024-3-19)}, 24,438 PRs\footnote{\url{https://github.com/tensorflow/tensorflow/pulls?q=is\%3Apr} (accessed on 2024-3-19)}, and 38,877 issues\footnote{\url{https://github.com/tensorflow/tensorflow/issues?q=is\%3Aissue} (accessed on 2024-3-19)} till March 2024. Therefore, in this study, we chose the TensorFlow repository on GitHub as the main source to build our dataset.
Moreover, Keras is an independent repository on GitHub, which is closely related to TensorFlow. For example, in some versions of TensorFlow, Keras has been integrated inside TensorFlow as a sub-module (i.e., \texttt{tensorflow.keras}). Therefore, in addition to TensorFlow, we also included the Keras repository on GitHub when constructing the AssuEval dataset~\cite{yangpackage2023}.

(2) \textbf{Determine the scope of the data}. According to similar works (e.g., \cite{Zhang2018}\cite{Ruan2019}\cite{Han2020}) that collect and analyze data on GitHub, in this study, we focus on commits, PRs, and issues of the selected repositories. 
There are also other sources (e.g., code comments) that may contain assumptions in the selected repositories. In this study, our objective was not to analyze assumptions (e.g., their importance and distribution) in DL framework projects. Instead, we aim to find a good algorithm and model to automatically identify assumptions in DL framework projects, and we consider that commits, PRs, and issues are representative data sources.

(3) \textbf{Determine the data period}. Repositories on GitHub are iteratively updated. The data period to be collected is from the creation of the TensorFlow and Keras repositories to the date we started the data collection process (that is, 2023-07-19). Data after this date are not included in this study.

(4) \textbf{Collect the data}. We used Assumption Miner~\cite{Yangtool2023} to collect the data. This tool automatically collected data from the GitHub sever using the GraphQL API\footnote{\url{https://docs.github.com/en/graphql/overview/about-the-graphql-api}}.

(5) \textbf{Review the collected data}. The first author verified the correctness and completeness of the collected data. For example, the first author visited the TensorFlow and Keras repositories on GitHub to check if the numbers of commits, PRs, and issues are consistent with the numbers in the local database.

% \begin{figure*} [!h]
%  \centering
%   \includegraphics [scale=0.13] {Figures/data_collection_process.png}
%  \caption{Process of data collection}
%  \label{data_collection}
% \end{figure*}

\subsection{Dataset management} \label{Dataset management}
\subsubsection{Construction process of the AssuEval dataset} \label{process}
\begin{figure*} [!h]
 \centering
  \includegraphics [scale=0.12] {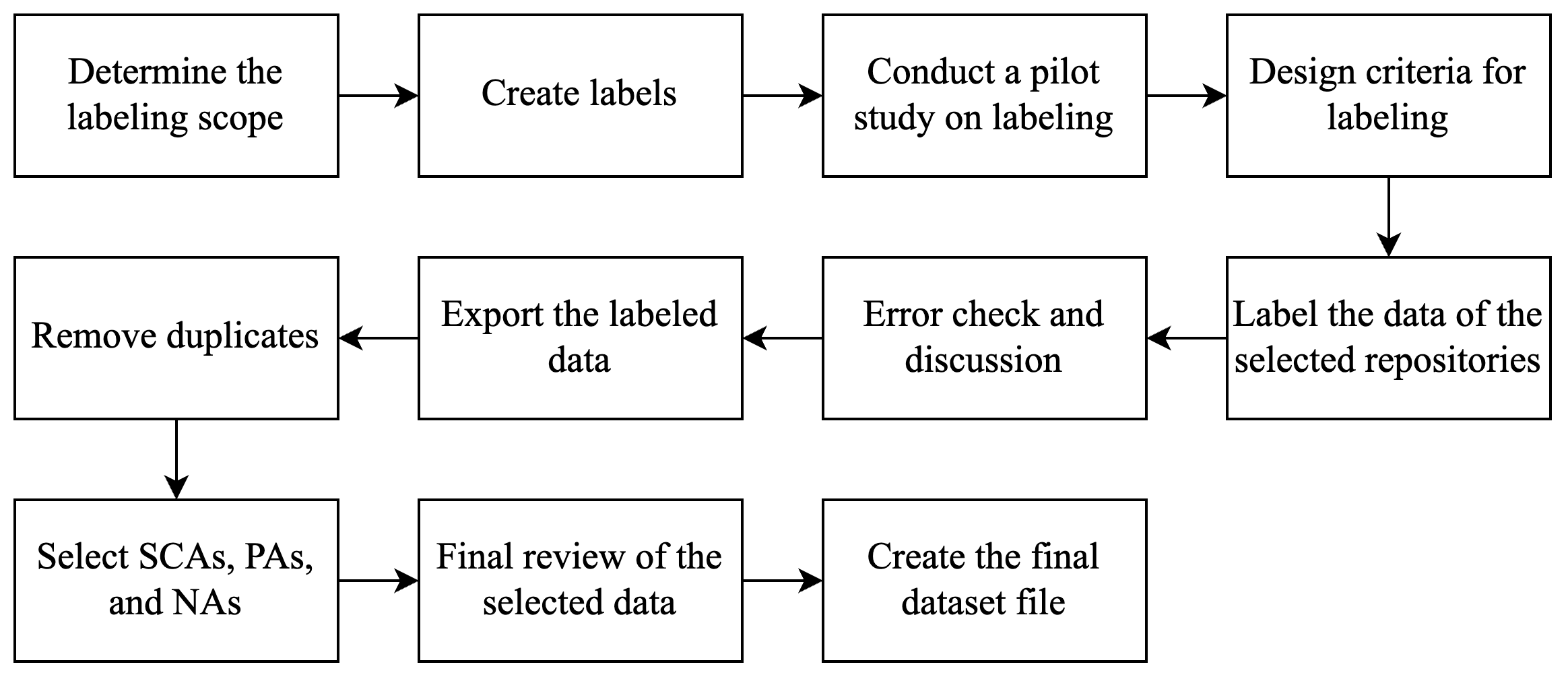}
 \caption{Construction process of the AssuEval dataset}
 \label{dataset}
\end{figure*}

The process of building the dataset is shown in Figure \ref{dataset}, which contains 11 steps as follows.

(1) \textbf{Determine the labeling scope}. Not all of the information in commits, PRs, and issues is meaningful for the identification of assumptions. Therefore, after analyzing the data models provided by GitHub, the scope of the data to be collected is shown in Table \ref{Scope of the data to be collected}. Details of the GraphQL API and the objects can be found in GitHub Docs\footnote{\url{https://docs.github.com/en/graphql/reference/objects}}.

\begin{table}[!h]
\scriptsize
\centering
\caption{Scope of the GitHub data to be collected}
\label{Scope of the data to be collected}
\begin{tabular}{ccc}
\toprule
\textbf{Object} & \textbf{Field} & \textbf{Description} \\
\midrule
Commit & message & The Git commit message. \\
PR & title & The title of the PR. \\
PR & body & The body of the PR. \\
PR & comments.body & A list of comments associated with the PR. \\
Issue & title & The title of the issue. \\
Issue & body & The body of the issue. \\
Issue & comments.body & A list of comments associated with the issue. \\
\bottomrule
\end{tabular}
\end{table}

(2) \textbf{Create labels}. We created three labels, i.e., NA with a value 0, PA with a value 1, and SCA with a value 2, which represent non-assumptions, potential assumptions, and self-claimed assumptions, respectively.

(3) \textbf{Conduct a pilot labeling}. We conducted a pilot labeling on the TensorFlow and Keras repositories. The criteria were refined iteratively during the pilot study. For example, regarding an assumption that is a condition (e.g., ``\textit{This obviously works if you can assume 0.0 is not real value in your data}''), we decided not to label such data as SCA, because the assumption does not exist. 
Another example is that if a sentence is a question (e.g., ```\textit{@willnorris for corporate CLAs, do we expect googlebot to say this has been signed? or do we just assume that if the user has @dropbox.com in their email that it is fine?}''), which is about whether to make certain assumptions, we decided to label such data as NA, since the assumption does not exist.
The third example is that if a sentence is a question (e.g., ``\textit{@yaroslavvb I’m assuming that if I run the probe op in a session together with computation of a model this would return me the peek memory usage, is that correct?}''), which is about the validity of an assumption rather than whether to make an assumption, we decided to label such data as SCA. 
More details of the included or excluded examples can be found in Table \ref{Inclusion criteria for SCAs}, Table \ref{Exclusion criteria for SCAs}, Table \ref{Inclusion criteria for PAs}, and Table \ref{Exclusion criteria for PAs}.

(4) \textbf{Design criteria for labeling}. After the pilot study, we formed the final criteria for labeling the data as mentioned in Section \ref{criteria}.

(5) \textbf{Label the data from the selected repositories}. The first author manually labeled the data following the criteria using Assumption Miner. NAs were labeled as ``0'', PAs were labeled as ``1'', and SCAs were labeled as ``2''.

(6) \textbf{Error check and discussion}. The first author checked the labeled data and then asked the other authors to review the results. All authors reached a consensus on the results.

(7) \textbf{Export the labeled data}. This step was also done by using Assumption Miner.

(8) \textbf{Remove duplicates}. Since commits, PRs, and issues may share some information, and the TensorFlow repository and the Keras repository are partly overlapped (e.g., TensorFlow uses Keras functions as high-level APIs), there could be duplicates in the labeled data. In this step, we removed all duplicates of the labeled data. Note that duplicates in this study mean that sentences are exactly the same, not only the meaning of sentences, but also the words, symbols, and spaces used. For example, we considered ``\# 32 floats \-> compression factor 24.5, assuming the input is 784 floats'' and ``\# 32 floats \-> compression of factor 24.5, assuming the input is 784 floats'' as two different data, though they have the same meaning, which can increase the diversity of the dataset.

(9) \textbf{Select SCAs, PAs, and NAs}. According to our previous studies (e.g., \cite{Yangsca2021} and \cite{Yangtool2023}), there are more than 1 million sentences in the TensorFlow and Keras repositories. The number of NAs is much higher than that of PAs, and the number of PAs is much higher than that of SCAs. Therefore, in this study, we selected all manually identified SCAs from the TensorFlow and Keras repositories. Then we randomly selected PAs and NAs with the same number of SCAs to construct the AssuEval dataset.

(10) \textbf{Final review of the selected data}. All authors conducted a final review on the selected data and reached a consensus on the labeling results.

(11) \textbf{Create the final dataset file}. After review, we created the final SCA, PA, and NA dataset file, which is available at \cite{yangpackage2023}. In the dataset file, the label ``0'' represents NA, ``1'' represents PA, and ``2'' represents SCA.

\subsubsection{Criteria of data labeling} \label{criteria}
\textbf{Motivation of defining criteria for the identification of assumptions}. Although stakeholders constantly make assumptions in their work \cite{Yangsms2018}\cite{Yangsurvey2016}, the definition or characteristics (e.g., subjective) of an assumption are abstract (e.g., something is uncertain but accepted as true without evidence) \cite{Yangsms2018}\cite{Yangtool2021}. As an example, Roeller et al. claimed that ``\textit{from one perspective or stakeholder, we may denote something as an assumption, while the same thing may be seen as a design decision from another perspective}''~\cite{Roeller2006}. This leads to difficulties in designing criteria to identify whether a sentence is an assumption in projects. In traditional assumption identification, most related studies chose to ask directly stakeholders if something is an assumption through experiments, surveys, and case studies. For example, as mentioned in the related work, Landuyt and Joosen collected 845 assumptions with 122 master students using a descriptive study \cite{Landuyt2020}. Such approaches require high costs, and the number of assumptions identified is usually limited ~\cite{Yangsms2018}. Since we need a large dataset, these approaches are not suitable for this study. 

\textbf{Inclusion and exclusion criteria for SCAs and PAs}. Based on the definition of the SCA concept and the PA concept, we designed the inclusion and exclusion criteria for SCAs with examples as shown in Table \ref{Inclusion criteria for SCAs} and Table \ref{Exclusion criteria for SCAs}, and the inclusion and exclusion criteria for PAs with examples as shown in Table \ref{Inclusion criteria for PAs} and Table \ref{Exclusion criteria for PAs}. Since the sentences in the commits, PRs, and issues are written by stakeholders of open source software (OSS) projects, the sentences may not be standard English. For example, considering the sentence: ``\textit{Perhaps @farizrahman4u's implementation could be used instead?}'', though it has a question mark, it is not the standard form of a question, and therefore we considered such sentences as not a question. For conditions, a sentence usually has two parts: a condition and a statement that is constrained by the condition. 
Considering that a sentence has an SCA, if the SCA is in a condition, we labeled it as NA, since it does not exist in the project, while if the SCA is in the statement that is constrained by a condition, we labeled it as SCA.
For alternatives, if making an SCA is only an option, we labeled it as NA (see examples in Table \ref{Exclusion criteria for SCAs}). This may be mixed with other types of alternatives. For example, we treat ``\textit{Not sure if it is necessary, but a more complicated alternative would be to assume that int size should mirror floatX.}'' as an SCA, since the alternative is regarding a solution to a problem, rather than whether to make an assumption.
These rules also apply to the inclusion and exclusion criteria for PAs. 
In this study, we considered assumption identification as a multiclass classification problem, and therefore we provided each sentence with only one label, that is, the labels are mutually exclusive. However, a sentence may be long and includes both SCAs and PAs (e.g., part of the sentence claims an SCA while the other part of the sentence describes a PA), which leads to the multiclass classification problem being a multi-label classification problem. We considered this as future work.

\scriptsize
\begin{longtable}{p{0.42\columnwidth}p{0.52\columnwidth}}
\caption{Inclusion criteria for self-claimed assumptions}
\label{Inclusion criteria for SCAs}\\
\hline
\toprule
\textbf{Inclusion criteria for SCAs} & \textbf{Examples} \\
\midrule
\multirow{3}{0.42\columnwidth}{\textbf{SCA\_IC1}: A sentence (a) contains valid or invalid assumptions using assumption-related keywords (including ``assumption'', ``assumptions'', ``assume'', ``assumes'', ``assumed'', ``assuming'', ``assumable'', and ``assumably'')  and (b) the assumptions have been made, i.e., not questions, conditions, and alternatives.} & {\textbf{E1}: \textit{For Ubuntu 21.04 the manual installation instructions are a thing of the past assuming that you have installed your NVIDIA drivers from the Canonical default apt repo.}} \\
& {\textbf{E2}: \textit{The main assumption behind the Glorot initialization is that the variance of the gradients should be the same in each layer.}} \\
& {\textbf{E3}: \textit{**Behavior**: When `tf.image.flip\_left\_right' is applied to this `tf.Tensor', the function incorrectly assumes a rank-3 shape and flips the image along the wrong axis.}} \\
\midrule
\multirow{3}{0.42\columnwidth}{\textbf{SCA\_IC2}: A sentence (a) indicates that there are valid or invalid assumptions (without details) somewhere using assumption-related keywords and (b) the assumptions have been made, i.e., not questions, conditions, and alternatives.} & {\textbf{E1}: \textit{\# Check input assumptions set after layer building, e.g. input shape.}} \\
& {\textbf{E2}: \textit{We tried to add lazy caching of the alternative row-partitioning tensors, but had to roll it back because it breaks some assumptions made by tf.cond.}} \\
& {\textbf{E3}: \textit{[XLA] Fix invalid assumption in HloComputation::CloneWithReplacements.}} \\
\midrule
\multirow{3}{0.42\columnwidth}{\textbf{SCA\_IC3}: A sentence (a) contains assumptions or indicates that there are assumptions (without details) somewhere using assumption-related keywords, (b) the assumptions are not conditions and alternatives, but they are in a question, and (c) the question does not concern whether something is an assumption or making certain assumptions, that is, such assumptions exist.} & {\textbf{E1}: \textit{Assuming that it is related, is there any caching involved when reading the .tfrecord?}} \\
& {\textbf{E2}: \textit{Can you update us on when all these features will be available and in which tensorflow version (only 2.0 I assume)?}} \\
& {\textbf{E3}: \textit{@yaroslavvb I'm assuming that if I run the probe op in a session together with computation of a model this would return me the peek memory usage, is that correct?}} \\
\midrule
\multirow{3}{0.42\columnwidth}{\textbf{SCA\_IC4}: A sentence refers to warnings that you should not make certain assumptions using assumption-related keywords.} & {\textbf{E1}: \textit{This CL generalizes the expansion such that it does not assume that `prefix' is a constant - it works for both constants and non-constants.}} \\
& {\textbf{E2}: \textit{Do not assume Node.in\_edges() is sorted by dst\_input.}} \\
& {\textbf{E3}: \textit{Update the parser test to check for unique'd hoisted map to be present but without assuming any particular order.}} \\
\bottomrule
\end{longtable}

\begin{longtable}{p{0.4\columnwidth}p{0.54\columnwidth}}
\caption{Exclusion criteria for self-claimed assumptions}
\label{Exclusion criteria for SCAs}\\
\hline
\toprule
\textbf{Exclusion criteria for SCAs} & \textbf{Examples} \\
\midrule
\multirow{3}{0.4\columnwidth}{\textbf{SCA\_EC1}: A sentence does not include any assumption-related keywords.} & {\textbf{E1}: \textit{Cast image preprocessing inputs to compute dtype.}} \\
& {\textbf{E2}: \textit{The improved version converts to a dynamic compatible form while restricting to static operations as often as possible.}} \\
& {\textbf{E3}: \textit{The problem occurs because PyYAML can't recognize numpy's data types.}} \\
\midrule
\multirow{3}{0.4\columnwidth}{\textbf{SCA\_EC2}: A sentence includes assumption-related keywords, and the keywords are used as the name of folders, files, classes, functions, variables, etc.} & {\textbf{E1}: \textit{old \= atomicCAS(address\_as\_ull, assumed, 0059 \_double\_as\_longlong(val + 0060 longlong\_as\_double(assumed)))}} \\
& {\textbf{E2}: \textit{ga\_uint old, assumed, sum, new}} \\
& {\textbf{E3}: \textit{Update '( assuming' to '(assuming'.}} \\
\midrule
\multirow{3}{0.4\columnwidth}{\textbf{SCA\_EC3}: A sentence includes assumption-related keywords, and assumptions are conditions or alternatives.} & {\textbf{E1}: \textit{If they are independent, or you make the assumption that they are, then you can do `-log(P(output1=1|data) * P(output2=c|data))' which can further decompose into `-log(P(output1=1|data))-log(P(output2=c|data))'.}} \\
& {\textbf{E2}: \textit{This obviously works if you can assume 0.0 is not real value in your data.}} \\
& {\textbf{E3}: \textit{So, if you assume independence between the two classifications, the linear combination of the two losses is indeed ``minimizing joint negative log likelihood''.}} \\
\midrule
\multirow{3}{0.4\columnwidth}{\textbf{SCA\_EC4}: A sentence includes assumption-related keywords, and the sentence is only about asking whether something is an assumption or whether to make certain assumptions, i.e., such assumptions do not exist.} & {\textbf{E1}: \textit{Are there any assumptions made by this method regarding the data?}} \\
& {\textbf{E2}: \textit{Can I assume that the parameter `y\_pred' in `custom\_objective(y\_true, y\_pred)' is just the list generated by my `generate\_batch\_data function' and I can treat it as a normal list?}} \\
& {\textbf{E3}: \textit{What is the assumed shape of 'weights'?}} \\
\bottomrule
\end{longtable}

\begin{longtable}{p{0.45\columnwidth}p{0.49\columnwidth}}
\caption{Inclusion criteria for potential assumptions}
\label{Inclusion criteria for PAs}\\
\hline
\toprule
\textbf{Inclusion criteria for PAs} & \textbf{Examples} \\
\midrule
\multirow{3}{0.45\columnwidth}{\textbf{PA\_IC1}: A sentence (a) is not an SCA, (b) contains valid or invalid expectations, future events, possibilities, guesses, opinions, feelings, or suspicions, and (c) the expectations, future events, etc. are not questions, conditions, and alternatives.} & {\textbf{E1}: \textit{If something/somewhere assumes that exceptions are available, things may not work as expected.}} \\
& {\textbf{E2}: \textit{Theano tile() expects Python int, so casting from numpy.int32 to Python int. (\#4330)}} \\
& {\textbf{E3}: \textit{My original docs only included the dimensions of the parameters (no batch dim) and were correct, but I think its better to change the functions to reflect the current docs.}} \\ \\ \\ \\
\midrule
\multirow{3}{0.45\columnwidth}{\textbf{PA\_IC2}: A sentence (a) is not an SCA, (b) contains valid or invalid expectations, future events, possibilities, guesses, opinions, feelings, or suspicions, (c) the expectations, future events, etc. are not conditions and alternatives, but are in a question, and (d) the question is not regarding whether to make expectations, guesses, opinions, feelings, or suspicions, whether future events are going to happen, or whether something is possible.} & {\textbf{E1}: \textit{Black image, seems a big wast to me, can I input another shape tensor instead?}} \\
& {\textbf{E2}: \textit{It would force people to clean up their existing code, and their code would get better. (Also, are there any concrete examples of existing code that assumes the current broken behavior?)}} \\
& {\textbf{E3}: \textit{an unpinned copy holds that mutex for a long time (until the copy completes?), potentially blocking other, unrelated work on the gpu.}} \\
\bottomrule
\end{longtable}

\begin{longtable}{p{0.45\columnwidth}p{0.49\columnwidth}}
\caption{Exclusion criteria for potential assumptions}
\label{Exclusion criteria for PAs}\\
\hline
\toprule
\textbf{Exclusion criteria for PAs} & \textbf{Examples} \\
\midrule
{\textbf{PA\_EC1}: A sentence is an SCA.} & {See the examples in Table \ref{Inclusion criteria for SCAs}.} \\
\midrule
\multirow{3}{0.45\columnwidth}{\textbf{PA\_EC2}: A sentence (a) includes expectations, future events, possibilities, guesses, opinions, feelings, or suspicions, while (b) they are conditions or alternatives.} & {\textbf{E1}: \textit{Allows fit() when call's signature looks something like call(x, training=true).}} \\
& {\textbf{E2}: \textit{Raise exception if unexpected keys are found in the padding dict.}} \\
& {\textbf{E3}: \textit{``It is diffcult or impossible to do obsure(very unique) things in Keras'', this is what an expert in out department has suggested me.}} \\
\midrule
\multirow{3}{0.45\columnwidth}{\textbf{PA\_EC3}: A sentence (a) includes expectations, future events, possibilities, guesses, opinions, feelings, or suspicions, while (b) the sentence consists of only asking whether to make expectations, guesses, opinions, feelings, or suspicions, whether future events are going to happen, or whether something is possible.} & {\textbf{E1}: \textit{hi @abattery, would you mind taking a look at this?}} \\
& {\textbf{E2}: \textit{it's not legal for collective-permtue to write to the same replica twice, because then which writer is supposed to win?}} \\
& {\textbf{E3}: \textit{make the community-builds link more prominent @andrew-leaver what do you think of this?}} \\
\midrule
\multirow{3}{0.45\columnwidth}{\textbf{PA\_EC4}: A sentence does not include any expectations, future events, possibilities, guesses, opinions, feelings, or suspicions.} & {\textbf{E1}: \textit{Added support for padded\_batch and fixed comments.}} \\
& {\textbf{E2}: \textit{Fix TypeError positional argument when used conjointly with tf-addons wrappers.}} \\
& {\textbf{E3}: \textit{* When use tensorflow as backend, let batch norm run into fused batch norm as much as possible, which has better performance.}} \\
\bottomrule
\end{longtable}
\normalsize

\subsection{Overview of the AssuEval dataset}
\begin{figure}[!htbp]     
\centering    
\includegraphics[scale=0.53]{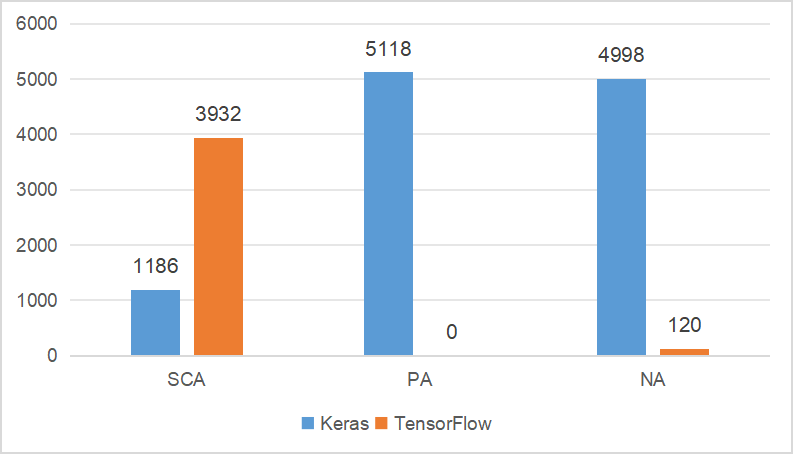}
\caption{Numbers of SCAs, PAs, and NAs in the AssuEval dataset selected from the TensorFlow and Keras repositories}
\label{number_repository}
\end{figure}

We collected 150,877 commits, 23,169 PRs, and 37,560 issues from the TensorFlow repository and 8,313 commits, 6,491 PRs, and 11,764 issues from the Keras repository. Based on the collected data, we further constructed the AssuEval dataset containing 5,118 SCAs (1,186, 23.17\% from Keras and 3,932, 76.83\% from TensorFlow), 5,118 PAs (100\% from Keras), and 5,118 NAs (4,998, 97.66\% from Keras and 120, 2.34\% from TensorFlow) as shown in Figure \ref{number_repository}. It is natural that different repositories may have different numbers of SCAs, PAs, and NAs, and we argue that the distribution of the data is not significant for this study. Researchers and practitioners can always use a subset of the dataset if they have specific requirements.

\begin{figure}[!htbp]   
\centering    
\includegraphics[scale=0.53]{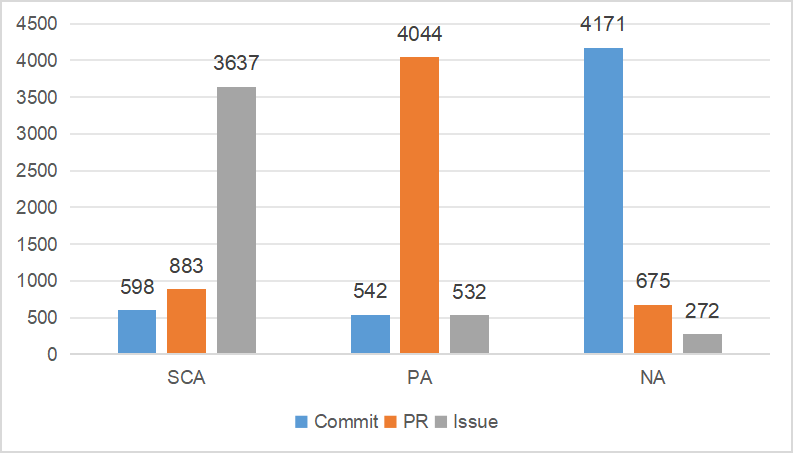}
\caption{Numbers of SCAs, PAs, and NAs in the AssuEval dataset across commits, PRs, and issues}
\label{number_commit_pr_issue}
\end{figure}

\begin{figure}[!htbp]     
\centering    
\includegraphics[scale=0.6]{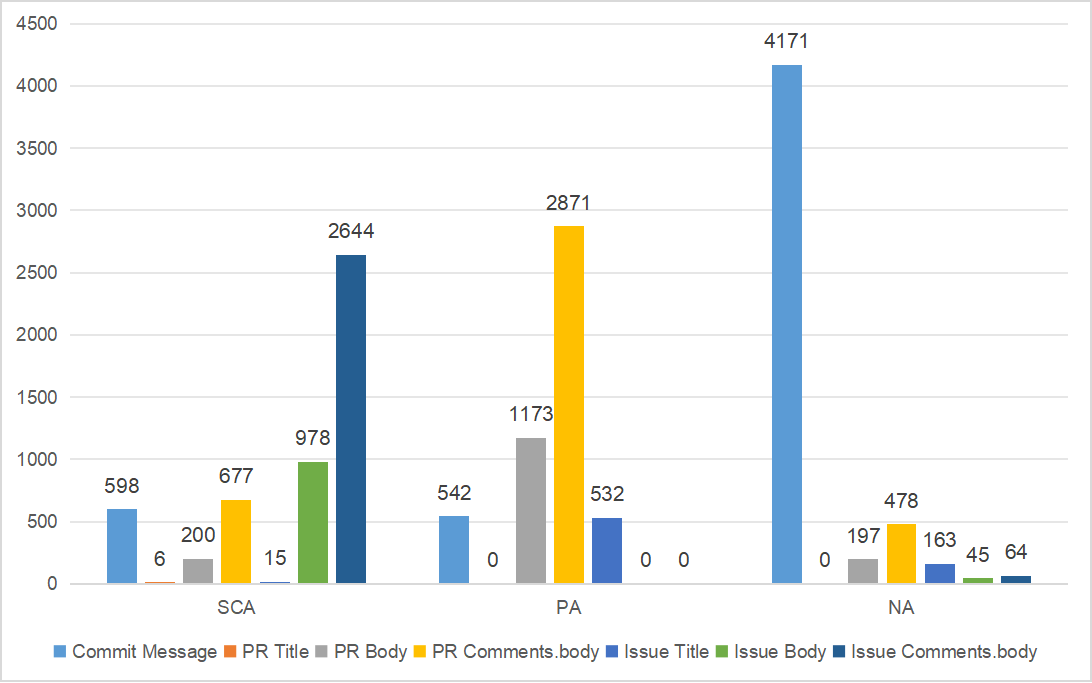}
\caption{Numbers of SCAs, PAs, and NAs in the AssuEval dataset across message of commits and title, body, and comments.body of PRs and issues}
\label{number_message_titile_body_comments}
\end{figure}

The sources of SCAs, PAs, and NAs include commits (598 SCAs, 542 PAs, and 4171 NAs), PRs (883 SCAs, 4044 PAs, and 675 NAs), and issues (3637 SCAs, 532 PAs, and 272 NAs), as shown in Figure \ref{number_commit_pr_issue}. Specifically, we collected data from (1) message of commits (598 SCAs, 542 PAs, and 4171 NAs), (2) title (6 SCAs), body (200 SCAs, 1173 PAs, and 197 NAs), and comments.body (677 SCAs, 2871 PAs, and 478 NAs) of PRs, and (3) title (15 SCAs, 532 PAs, and 163 NAs), body (978 SCAs and 45 NAs), and comments.body of issues (2644 SCAs and 64 NAs), as shown in Figure \ref{number_message_titile_body_comments}.

\begin{figure}[!htbp]     
\centering    
\includegraphics[scale=0.54]{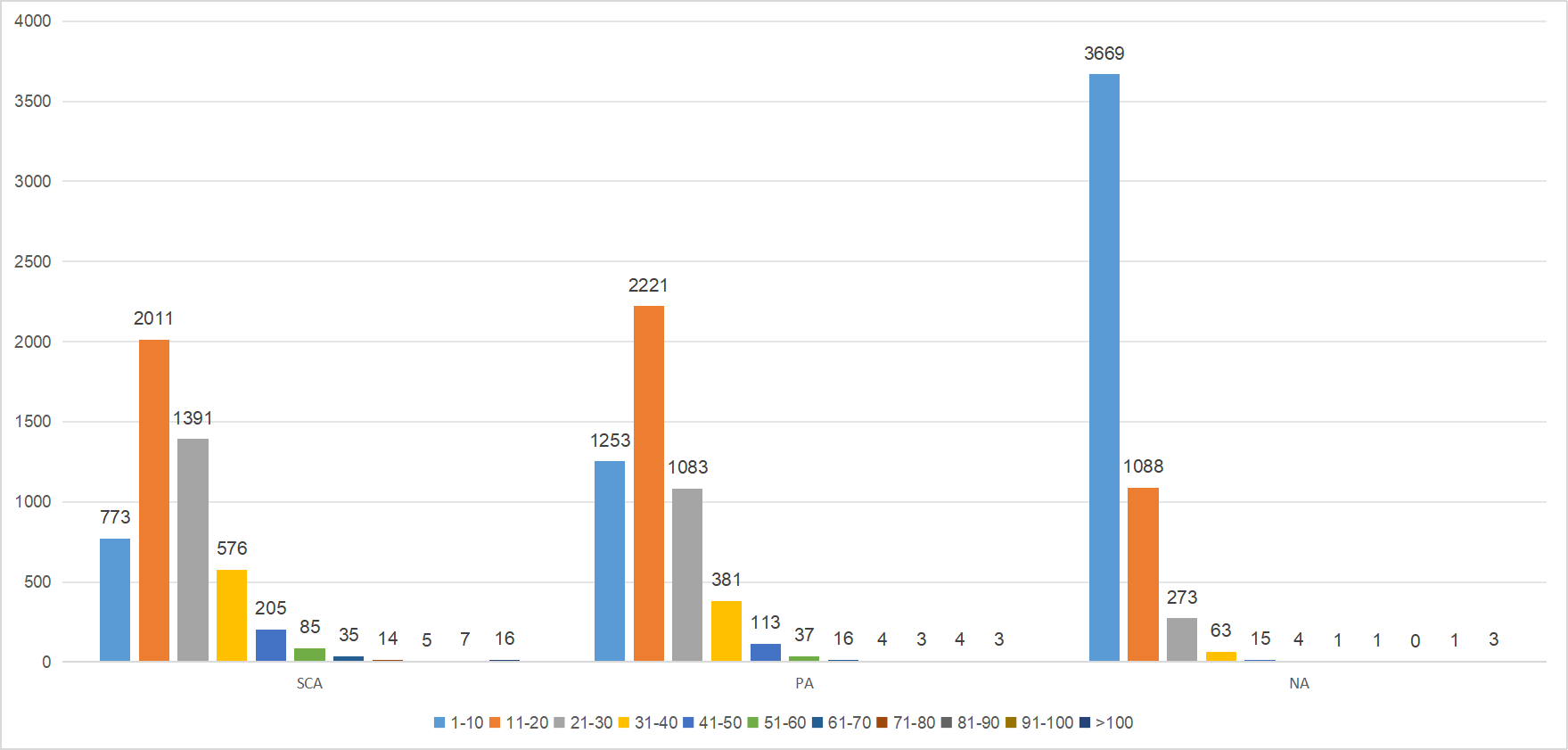}
\caption{Summary of the numbers of words in each data item in the AssuEval dataset}
\label{number_of_words}
\end{figure}

We also counted the number of words for each data item (punctuation masks in a sentence were not counted). The numbers of words for each SCA, PA, and NA range from 2 to 276, 2 to 102, and 1 to 180, respectively. To keep the figure short, we arranged the numbers into 11 intervals, as shown in Figure \ref{number_of_words}. The length of the first ten intervals is 10 (e.g., 1-10 means the sum of data items with a number of words between 1 and 10). In the figure, we can see that most of the SCAs (4,956 out of 5,118, 96.83\%), PAs (5,051 out of 5,118, 98.69\%), and NAs (5,108 out of 5,118, 99.80\%) have less than 50 words.

\section {Study design} \label{design}
The objective of the study is to evaluate different classification models for the purpose of identification with respect to assumptions from the point of view of developers and users in the context of DL framework projects (i.e., issues, pull requests, and commits) on GitHub. 
As defined by Wohlin and Rainer \cite{Wohlin2022}: ``\textit{A case study is an empirical investigation of a case, using multiple data collection methods, to study a contemporary phenomenon in its real-life context, and with the investigator(s) not taking an active role in the case investigated}.'' The cases in our study are the two DL framework projects: TensorFlow and Keras, which are hosted on GitHub and represent real-life phenomena in the software engineering domain. The data collected was from their real-life context; all the issues, pull requests, and commits were part of the natural development workflow of these two projects and documented by their stakeholders. While we utilized the data to analyze the performance of classification models, we did so without altering or influencing the projects themselves. Our role in this study was observational and analytical. We recognize the definition of a case study provided by Wohlin and Rainer \cite{Wohlin2022} and acknowledge that there are some distinctions between our work and the definition in \cite{Wohlin2022}. However, we believe that the core principles of case studies are applicable to our study. Therefore, instead of setting hypotheses, which are often employed in experiments, we followed the guidelines \cite{Runeson2012} proposed by Runeson \textit{et al.} to design and conduct the study.

In this study, we used and analyzed seven non-transformers based models (e.g., Support Vector Machine, Classification and Regression Trees), the ALBERT model, and three decoder-only models (i.e., ChatGPT, Claude, and Gemini). 
The selection of models could be different in other studies. However, our selection is either from related work of assumptions in software development or popular models frequently mentioned in the literature. Therefore, we believe that the models selected in this study are representative and the results can be generalized to other similar models.
% Moreover, for some cases, ALBERT may also be considered as an LLM. We did not
% train the ALBERT model with a large text corpus, and therefore we treated the ALBERT model as a
% DL model instead of an LLM.

Additionally, rule-based approaches may be suitable for identifying SCAs (i.e., searching ``assum'' in sentences), but we see the following flaws: (1) There are many variables, functions, etc. named with ``assum'' (e.g., ``assume''), which is difficult to filter out in all cases. (2) Sentences may have terms with ``assum'', while the sentences are about whether to make certain assumptions (e.g., ``\textit{@willnorris for corporate CLAs, do we expect googlebot to say this has been signed? or do we just assume that if the user has @dropbox.com in their email that it is fine?}''), that is, the assumptions do not exist. Using a simple rule-based approach may misclassify such cases as SCAs, which are NAs. (3) Rule-based approaches may not be suitable for the identification of PAs based on the definition. For example, consider the following two sentences that have the term ``possible'': a. ``\textit{Along with the template, please provide as many details as possible to find the root cause of the issue like CUDA and cuDNN versions.}'' and b. ``\textit{@YANxu-666: we don't have any continuous integration for Windows so it is possible that something in the script is not Windows-compatible.}'', Sentence \textit{a} is clearly not an assumption while Sentence \textit{b} is a PA. In such a case, considering phrases instead of words may address the problem. However, there are many words that can lead to assumptions, and the phrases can add much more complexity to the assumption identification algorithm.
Therefore, in this study, we did not analyze the performance of rule-based approaches in identifying assumptions.

\subsection{Research questions}
\textbf{RQ1: Which non-transformers based model achieves the best performance on the AssuEval dataset?}

\textbf{Rationale}: Algorithms usually have varying performance in classification problems, leading to different trained models. For example, Naive Bayes models are a group of probabilistic classifiers that have been widely used in text categorization since 1960. The Support Vector Machine algorithm is a commonly used supervised ML algorithm for data classification and regression analysis. 
To answer RQ1, we chose seven popular non-transformers based algorithms to train classification models for the identification of assumptions: Perceptron (Pct), Logistic Regression (LR), Linear Discriminant Analysis (LDA), K-Nearest Neighbors (KNN), Support Vector Machine (SVM), Naive Bayes (NB), and Classification and Regression Trees (CART). These algorithms have been used in similar studies, such as the classification of non-functional requirements \cite{Abad2017} and rationale \cite{Alkadhi2018}. Li \textit{et al.} used these seven non-transformers based algorithms to identify assumptions (binary classification) in the Hibernate project \cite{Li2019}. Compared to their work, the scale of the dataset used in this study is 19 times larger (15,384 vs. 800).
Through this RQ, we plan to evaluate the performance of non-transformers based models to identify assumptions on the AssuEval dataset \cite{yangpackage2023}.

\textbf{RQ2: What is the performance of ALBERT on the AssuEval dataset?}

\textbf{Rationale}: In addition to non-transformers based models, transformers based models have also been widely used in various areas as a key component, such as image classification, question-and-answer, and speech recognition. 
For answering RQ2, we chose and trained a representative encoder-only model (i.e., ALBERT) to identify assumptions. ALBERT is widely used in many similar tasks, such as the classification of sentiments in Yelp reviews \cite{Alamoudi2021}, the classification of Japanese text \cite{Zhang2022}, and the detection of specific speech in a given sentence \cite{Vijayakumar2022}.
The results of this RQ provide a comparison of the performance in identifying assumptions between the ALBERT model and the non-transformers based models.

\textbf{RQ3: Are decoder-only models able to perform the task of identifying assumptions?}

\textbf{Rationale}: Besides the selected models in RQ1 and RQ2, decoder-only models are another potential solution to identify assumptions. 
In this RQ, we chose three of the most representative decoder-only models (i.e., ChatGPT, Claude, and Gemini) to explore their performance in identifying assumptions. ChatGPT, Claude, and Gemini are advanced AI chatbots developed by OpenAI\footnote{\url{https://www.openai.com}}, Anthropic\footnote{\url{https://www.anthropic.com}}, and Google\footnote{\url{https://www.google.com}}, which can interact in a conversational way \cite{chatgptintro2022}. As reported by OpenAI, more than 100 million people used ChatGPT. These chatbots can be applied in various fields, such as answering specific questions, writing an essay, writing a song, writing code to implement a function, and code debugging \cite{chatgptintro2022}\cite{Meyer2023}. 
Specifically, we chose ChatGPT (based on the GPT-3.5 model and the GPT-4 model), Claude (based on the Claude 3.5 Sonnet model), and Gemini (based on the Gemini 1.0 pro model). 
The results of this RQ show the performance of the decoder-only models in identifying assumptions on the AssuEval dataset.

\subsection{Training}
We deployed Assumption Miner locally to train the non-transformers based models and the ALBERT model. The configuration of the computer we used for the training is: (1) CPU: 12th Gen Intel(R) Core(TM) i7-12700F, (2) Memory: 32GB, (3) Hard Disk: 500GB SSD, (4) GPU: NVIDIA GeForce RTX 4060 Ti.

\subsubsection{Data processing}
We first split the dataset into a training set (80\%) and a testing set (20\%), which is a common strategy used in many classification-related tasks (e.g., \cite{Jasim2021}\cite{Helber2019}\cite{Durden2021}). When loading the data, we transformed all the loaded data into the lower case. Then we used the same vocabulary with a size of 30,000 as in ALBERT \cite{LanALBERT2020}, and we used SentencePiece to tokenize the data in the training set and the test set \cite{Kudo2018}. The vocabulary (``30k-clean.vocab'' and ``30k-clean.model'') can be found in the replication package of this study \cite{yangpackage2023}.

\subsubsection{Model construction and training}
For answering RQ1, we used the scikit-learn package to implement and train the seven non-transformers based models. The configurations of each model are shown in Table \ref{Configurations of training the non-transformers based models}, which is based on the settings recommended in the scikit-learn package.

\begin{sidewaystable}[!h]
\scriptsize
\centering
\caption{Configurations of training the non-transformers based models}
\label{Configurations of training the non-transformers based models}
\begin{tabular}{cccccccc}
\toprule
\multicolumn{2}{c}{Pct} & \multicolumn{2}{c}{LR} & \multicolumn{2}{c}{SVM} & \multicolumn{2}{c}{CVAT} \\
\midrule
tol & 1e-3 & tol & 1e-4 & tol & 1e-3 & criterion & gini \\
penalty & None & penalty & l2 & coef0 & 0.0 & splitter & best \\
l1\_ratio & 0.15 & l1\_ratio & None & cache\_size & 200 & max\_depth & None \\
fit\_intercept & True & fit\_intercept & True & break\_ties & False & min\_samples\_split & 2 \\
max\_iter & 1,000 & max\_iter & 10,000 & max\_iter & -1 & min\_samples\_leaf & 1 \\
verbose & 0 & verbose & 0 & verbose & False & min\_weight\_fraction\_leaf & 0 \\
validation\_fraction & 0.1 & solver & lbfgs & kernel & rbf & max\_features & None \\
n\_jobs & None & n\_jobs & None & decision\_function\_shape & ovr & max\_leaf\_nodes & None \\
random\_state & 0 & random\_state & None & random\_state & None & random\_state & None \\
alpha & 0.0001 & dual & False & gamma & scale & min\_impurity\_decrease & 0 \\
class\_weight & None & class\_weight & None & class\_weight & None & class\_weight & None \\
eta0 & 1.0 & intercept\_scaling & 1 & degree & 3 & ccp\_alpha & 0.0 \\
early\_stopping & False & multi\_class & auto & probability & False & min\_impurity\_split & None \\
n\_iter\_no\_change & 5 & C & 1.0 & C & 1.0 & & \\
warm\_start & False & warm\_start & False & shrinking & True &  &  \\
shuffle & True &  &  &  &  &  &  \\
\midrule
\multicolumn{2}{c}{LDA} & \multicolumn{2}{c}{KNN} & \multicolumn{2}{c}{NB} & \multicolumn{2}{c}{}\\
\midrule
tol & 1e-4 & n\_neighbors & 5 & alpha & 1.0  &  & \\
shrinkage & None & weights & uniform & binarize & .0  &  & \\
priors & None & algorithm & auto & fit\_prior & True  &  & \\
n\_components & None & leaf\_size & 30 & class\_prior & None  &  & \\
store\_covariance & False & p & 2 &  &  &  & \\
covariance\_estimator & None & metric & minkowski &  &  &  & \\
solver & svd & metric\_params & None &  &  &  & \\
 &  & n\_jobs & None &  &  &  & \\
\bottomrule
\end{tabular}
\end{sidewaystable}

\clearpage

To answer RQ2, we used the TensorFlow framework to train the ALBERT model. The configurations of the ALBERT model are shown in Table \ref{Configurations of training the ALBERT model}, which is based on the recommendation by the Google team on GitHub\footnote{\url{https://github.com/google-research/albert}}. The optimizer is Adam, the training batch size is 64, and the maximum sequence length of each sentence is 128. We trained the model to 10,000 epochs with an initial learning rate of 2e-5.

\begin{table}[!h]
\scriptsize
\centering
\caption{Configurations of training the ALBERT model}
\label{Configurations of training the ALBERT model}
\begin{tabular}{cc}
\toprule
\textbf{Item} & \textbf{Value} \\
\midrule
attention\_probs\_dropout\_prob & 0 \\
hidden\_act & gelu \\
hidden\_dropout\_prob & 0 \\
embedding\_size & 128 \\
hidden\_size & 768 \\
initializer\_range & 0.02 \\
intermediate\_size & 3,072 \\
max\_position\_embeddings & 512 \\
num\_attention\_heads & 12 \\
num\_hidden\_layers & 12 \\
num\_hidden\_groups & 1 \\
net\_structure\_type & 0 \\
gap\_size & 0 \\
num\_memory\_blocks & 0 \\
inner\_group\_num & 1 \\
down\_scale\_factor & 1 \\
type\_vocab\_size & 2 \\
vocab\_size & 30,000 \\
\bottomrule
\end{tabular}
\end{table}

To answer RQ3, we used ChatGPT (based on the GPT-3.5 model and the GPT-4 model), Claude (based on the Claude 3.5 Sonnet model), and Gemini (based on the Gemini 1.0 pro model) without retraining or further fine-tuning. 
In this study, we first asked these models a set of warm-up questions (as shown in Table \ref{Questions for ChatGPT}) about assumptions and their identification in both general software development and DL framework development. Our objective was not to explore how the selected decoder-only models work in answering questions but to check whether the selected decoder-only models have a good understanding of this field. Subsequently, we provided the selected decoder-only models with the process and rules of the task, the definitions of SCA, PA, and NA (as mentioned in Section \ref{assuevaldataset}) and the criteria for identifying SCAs, PAs, and NAs mentioned in Section \ref{criteria}. Then we provided each sentence from the test set to the selected decoder-only models and asked them to classify it as SCA, PA, or NA. Details of the chat with the selected decoder-only models can be found in~\cite{yangpackage2023}.

\begin{table}[!h]
\scriptsize
\centering
\caption{Warm-up questions for the selected decoder-only models}
\label{Questions for ChatGPT}
\begin{tabular}{cp{0.9\columnwidth}}
\toprule
\textbf{No.} & \multicolumn{1}{c}{\textbf{Questions}} \\
\midrule
{1} & {What is an assumption in software development?} \\
{2} & {Are the assumptions different in general software development and deep learning framework development?} \\
{3} & {Could assumptions in deep learning framework development be related to other assumptions or other types of software artifacts such as requirements, design decisions, and bugs?} \\
{4} & {Is identifying assumptions important in deep learning framework development?} \\
{5} & {Many deep learning framework projects are on GitHub. What sources of a deep learning framework project on GitHub may contain assumptions?} \\
{6} & {How do you identify assumptions in the sources of a deep learning framework project on GitHub?} \\
\bottomrule
\end{tabular}
\end{table}

\subsection{Data analysis}
\begin{figure*} [!h]
 \centering
  \includegraphics [scale=0.07] {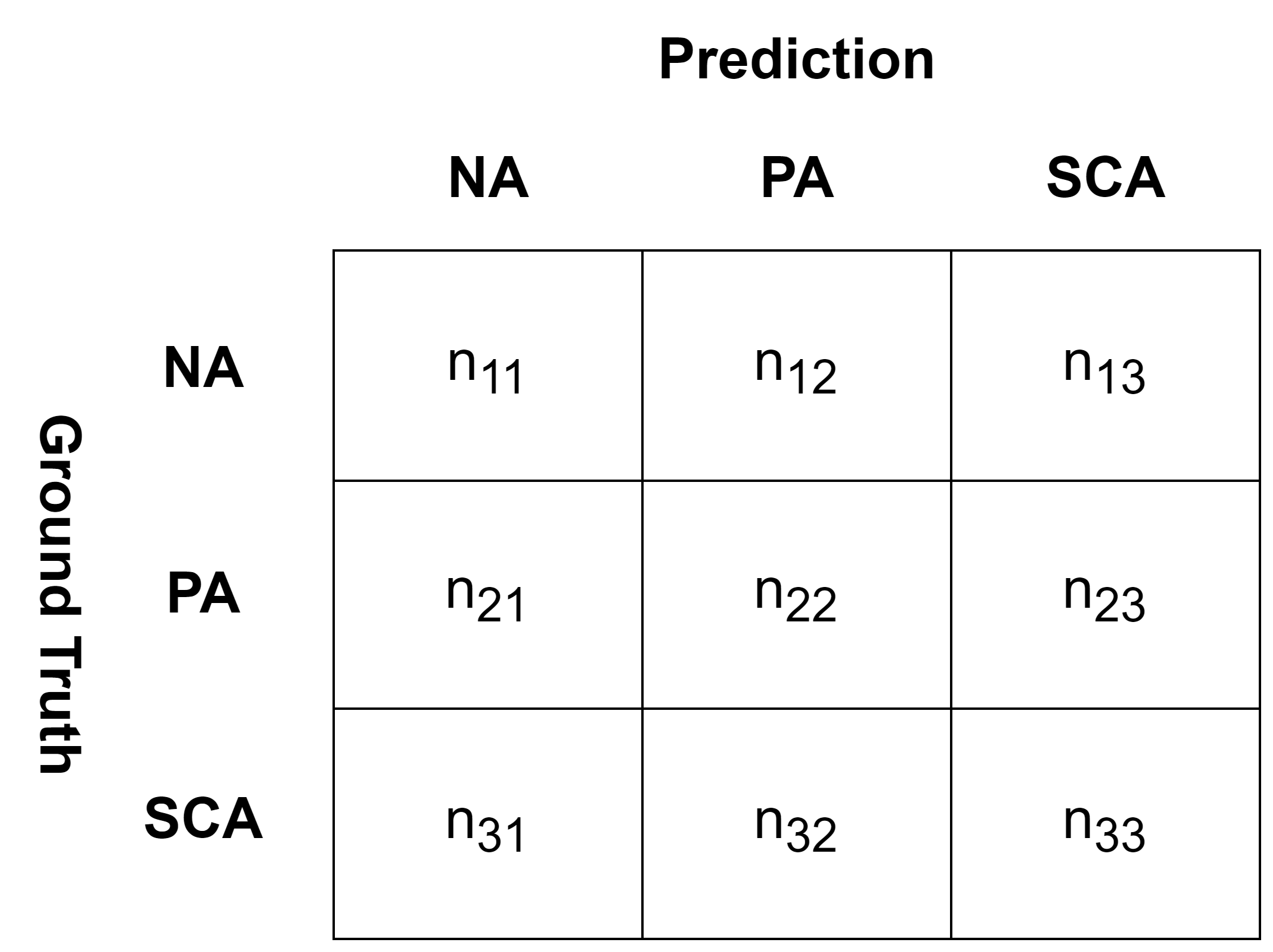}
 \caption{Confusion matrix of the AssuEval dataset}
 \label{confusion_matrix}
\end{figure*}

For a multiclass classification problem, the confusion matrix is an $n \times n$ matrix, where $n$ is the number of classes. Each row represents the instances of an actual class and each column represents the instances of a predicted class.
Since the AssuEval dataset (see Section \ref{Dataset management}) has three labels, the confusion matrix of the dataset is shown in Figure \ref{confusion_matrix}. $n_{ij} (i = 1, 2, 3, j = 1, 2, 3)$ represents the number of results with respect to different conditions. 

\begin{figure*} [!h]
 \centering
  \includegraphics [scale=0.07] {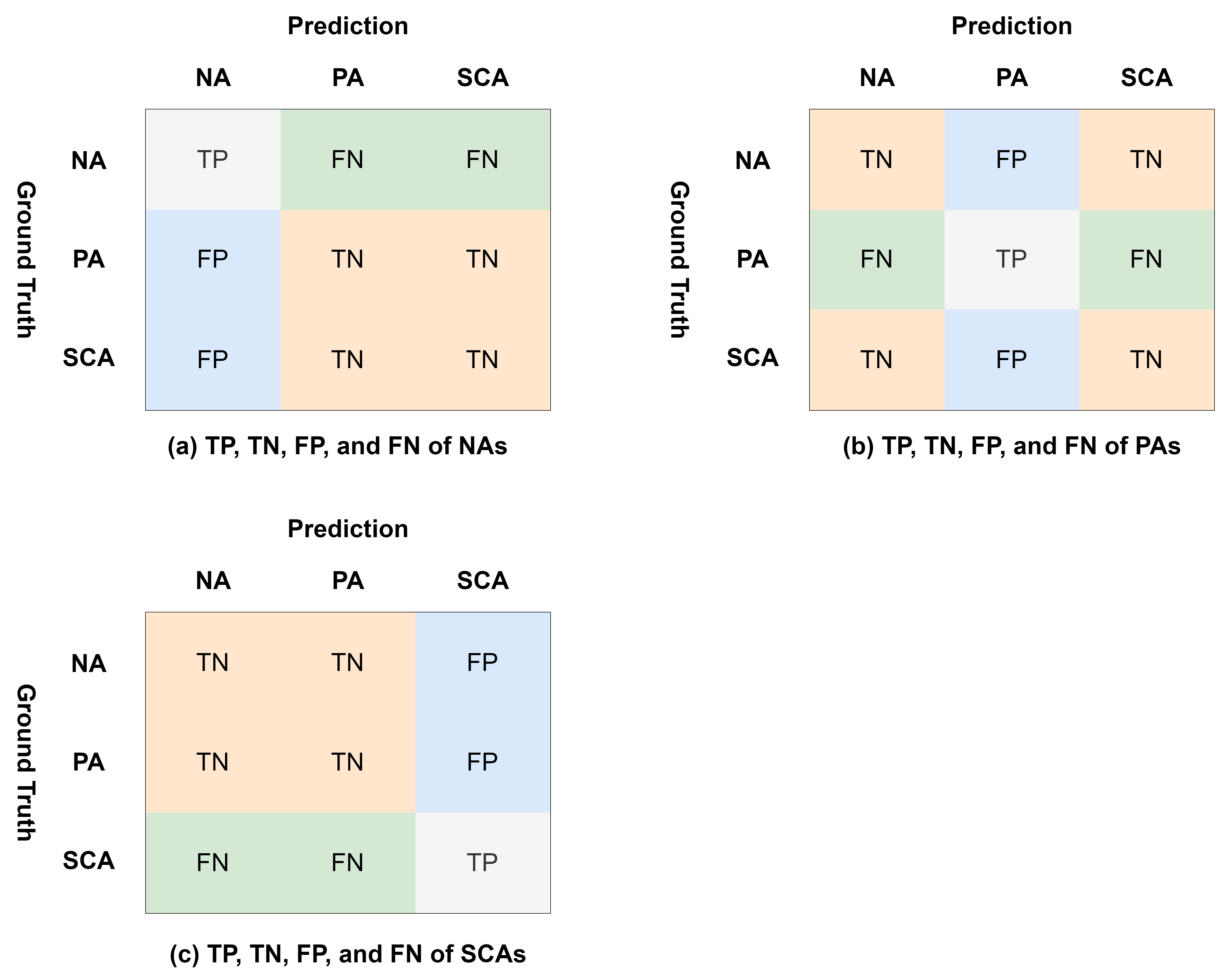}
 \caption{TP, TN, FP, and FN of each label}
 \label{TP_TN_FP_FN}
\end{figure*}

True positive (TP) is a number of positive labels that are correctly identified as positive.
True negative (TN) is a number of negative labels that are correctly identified as negative.
False positive (FP) is a number of negative labels that are incorrectly identified as positive.
False negative (FN) is a number of positive labels that are incorrectly identified as negative.
In a multiclass classification problem, the terms ``positive'' and ``negative'' are not used in the same way as in binary classification \cite{Sokolova2009}. Instead, each class is considered separately against all other classes \cite{Sokolova2009}. Therefore, in this case, when considering NA, TP represents the instances correctly predicted as NA. FP represents the instances that are not NA but are incorrectly predicted as NA. TN represents the instances that are not NA and are correctly predicted to be not NA. FN represents the instances that are NA but are incorrectly predicted to be not NA. The same logic applies when considering PA or SCA. Therefore, each class gets its turn to be the ``positive'' class, with all other classes combined to form the ``negative'' class. This allows us to measure the performance of the classifiers for each class versus all other classes.
The TP, TN, FP, and FN of each label are shown in Figure \ref{TP_TN_FP_FN}.

\begin{figure*} [!h]
 \centering
  \includegraphics [scale=0.07] {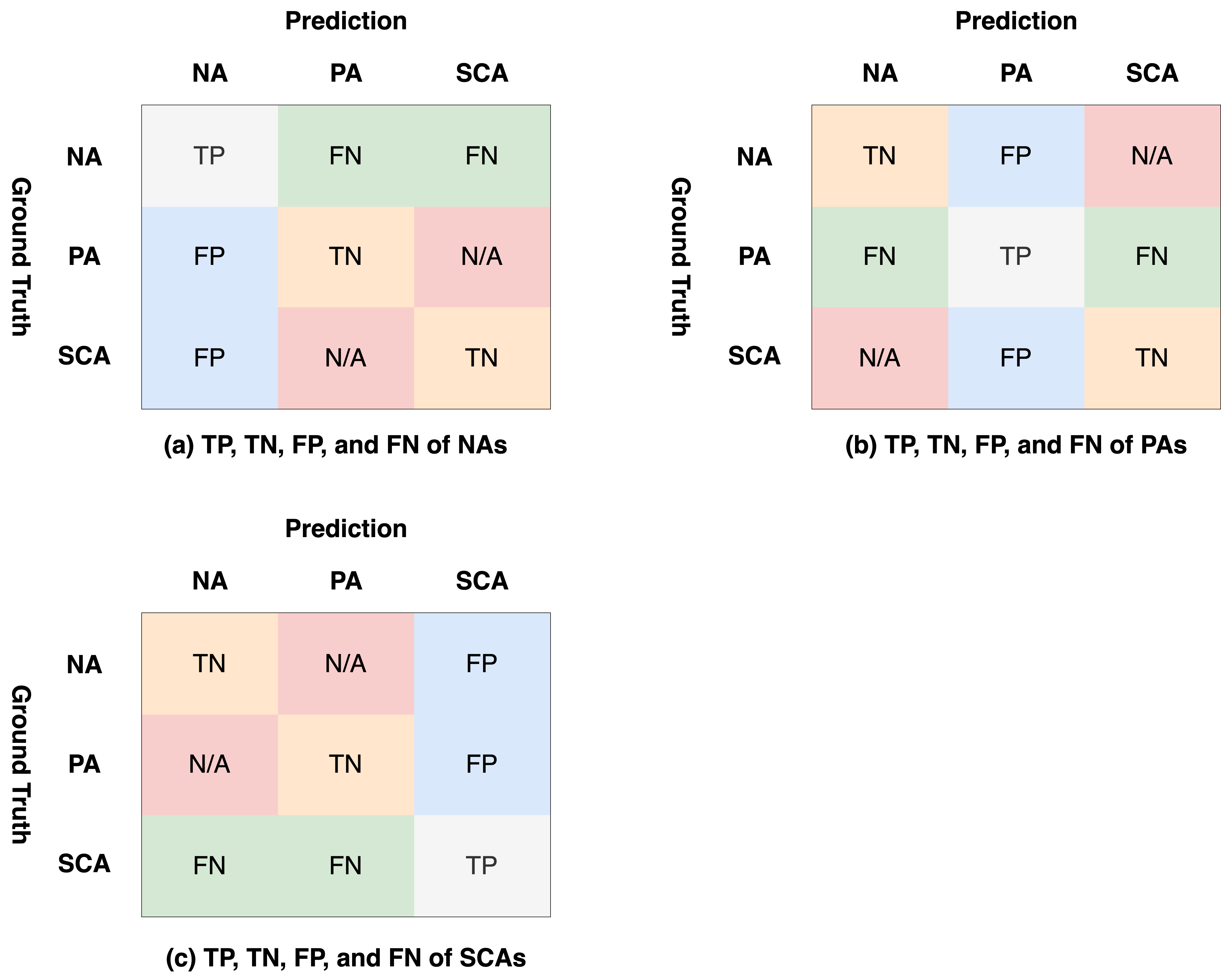}
 \caption{TP, TN, FP, and FN of each label}
 \label{TP_TN_FP_FN_2}
\end{figure*}

In this study, we adopted a more strict standard to calculate the classification performance, that is, for TN, the predicted label for a sample should match exactly the corresponding label on the ground truth, as shown in Figure \ref{TP_TN_FP_FN_2}. ``N/A'' means ``Not Applicable''.
For instance, when examining the confusion matrix of NAs, if a sample is labeled SCA and the model classifies the sample as PA, even though it accurately identified negative labels (both SCA and PA are negative labels in this case), the model prediction is incorrect (PA) compared to the ground truth (SCA). Hence, it should not be considered a ``True'' negative.

We used accuracy, precision, recall, and f1-score to measure the classification performance of different models, as shown in Equation \ref{eq15}. 

\begin{equation}\label{eq15}
\begin{split}
&Accuracy = \frac{TP + TN}{TP + FP + FN + TN} \\
&Precision=\frac{TP}{TP + FP} \\
&Recall=\frac{TP}{TP + FN} \\
&F1-score=\frac{2 * Precision * Recall}{Precision + Recall}
\end{split}
\end{equation}

Since the problem is multiclass classification and the dataset is balanced, we used a macro average strategy for precision and recall calculations \cite{Grandini2020}. Taking into account that the precision of SCAs, PAs, and NAs is $P_{SCA}$, $P_{PA}$, and $P_{NA}$ and that the recall of SCAs, PAs, and NAs is $R_{SCA}$, $R_{PA}$, and $R_{NA}$, then the precision and recall are shown in Equation \ref{eq16}.

\begin{equation}\label{eq16}
\begin{split}
&Precision_{macro}=\frac{P_{SCA} + P_{PA} + P_{NA}}{3} \\
&Recall_{macro}=\frac{R_{SCA} + R_{PA} + R_{NA}}{3} \\
\end{split}
\end{equation}

\section {Results} \label{results}
\subsection{Results of RQ1}
\begin{figure}[!htbp]     
\centering    
\includegraphics[scale=0.8]{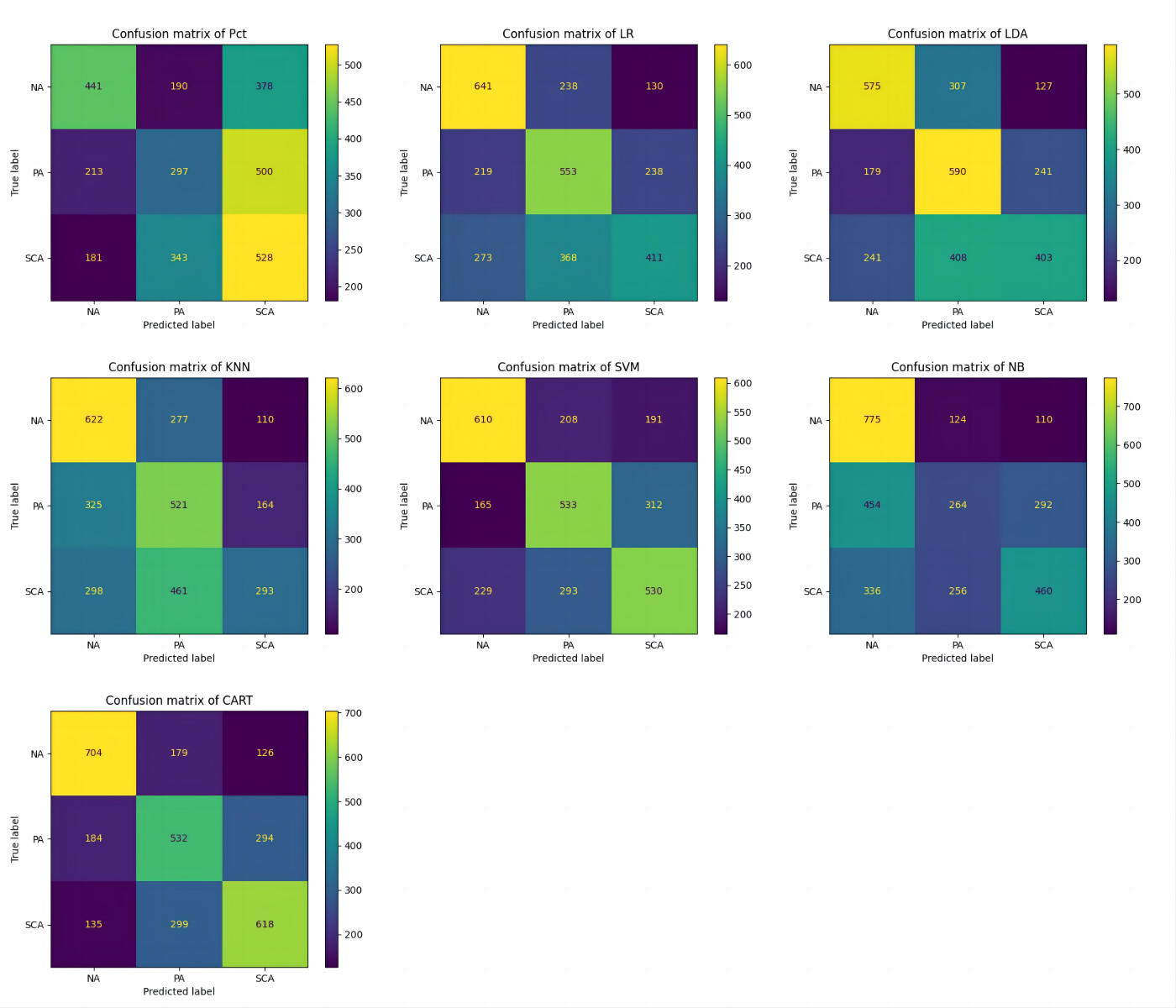}
\caption{Confusion matrices of the non-transformers based models}
\label{confusion_matrices}
\end{figure}

The confusion matrices of the Pct, LR, LDA, KNN, SVM, NB, and CART models are shown in Figure \ref{confusion_matrices}. The accuracy, precision, recall, and f1-score of each model are shown in Table \ref{Performance of the seven models}. 

\textbf{Summary}: The CART model achieves the best accuracy, precision, recall, and f1-score to identify SCAs, PAs, and NAs, while the worst performance in identifying SCAs, PAs, and NAs is the Pct model.

\begin{table}[!htbp]
\scriptsize
\centering
\caption{Performance of the seven non-transformers based models of assumption identification on the AssuEval dataset}
\label{Performance of the seven models}
\begin{tabular}{ccccc}
\toprule
Model & Accuracy & Precision & Recall & F1-score \\
\midrule
Pct & 0.4122 & 0.4205 & 0.4110 & 0.4103 \\
LR & 0.5226 & 0.5235 & 0.5245 & 0.5191 \\
LDA & 0.5106 & 0.5176 & 0.5124 & 0.5086 \\
KNN & 0.4676 & 0.4767 & 0.4703 & 0.4577 \\
SVM & 0.5448 & 0.5454 & 0.5454 & 0.5453 \\
NB & 0.4881 & 0.4796 & 0.4889 & 0.4674 \\
CART & \cellcolor{lightgray}0.6037 & \cellcolor{lightgray}0.6034 & \cellcolor{lightgray}0.6040 & \cellcolor{lightgray}0.6037 \\
\bottomrule
\end{tabular}
\end{table}

\subsection{Results of RQ2}
\begin{figure}[!htbp]     
\centering    
\includegraphics[scale=0.6]{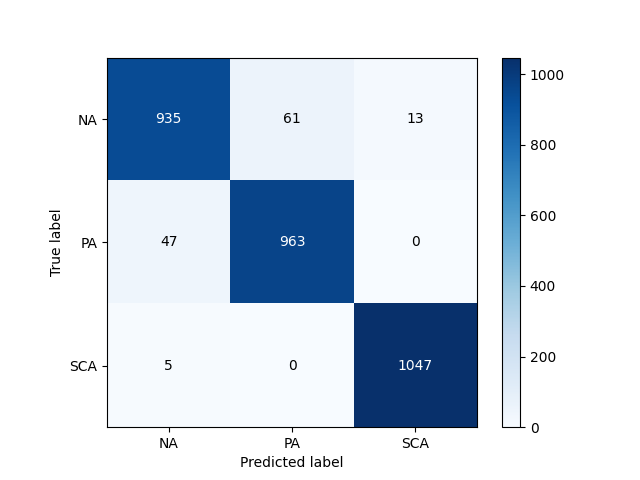}
\caption{Confusion matrices of the ALBERT model}
\label{confusion_matrix_albert}
\end{figure}

The confusion matrix of the ALBERT model is shown in Figure \ref{confusion_matrix_albert}. Compared to the best model (i.e., the CART model) in RQ1, although training the ALBERT model requires much more resources (e.g., seconds vs. hours for training), the trained ALBERT model performs much better than the seven non-transformers based models trained on the same dataset.

\textbf{Summary}: The accuracy, precision, recall, and f1-score of the ALBERT model are 0.9590, 0.9585, 0.9585, and 0.9584, respectively.

\subsection{Results of RQ3}
\begin{figure}[!htbp]     
\centering    
\includegraphics[scale=0.35]{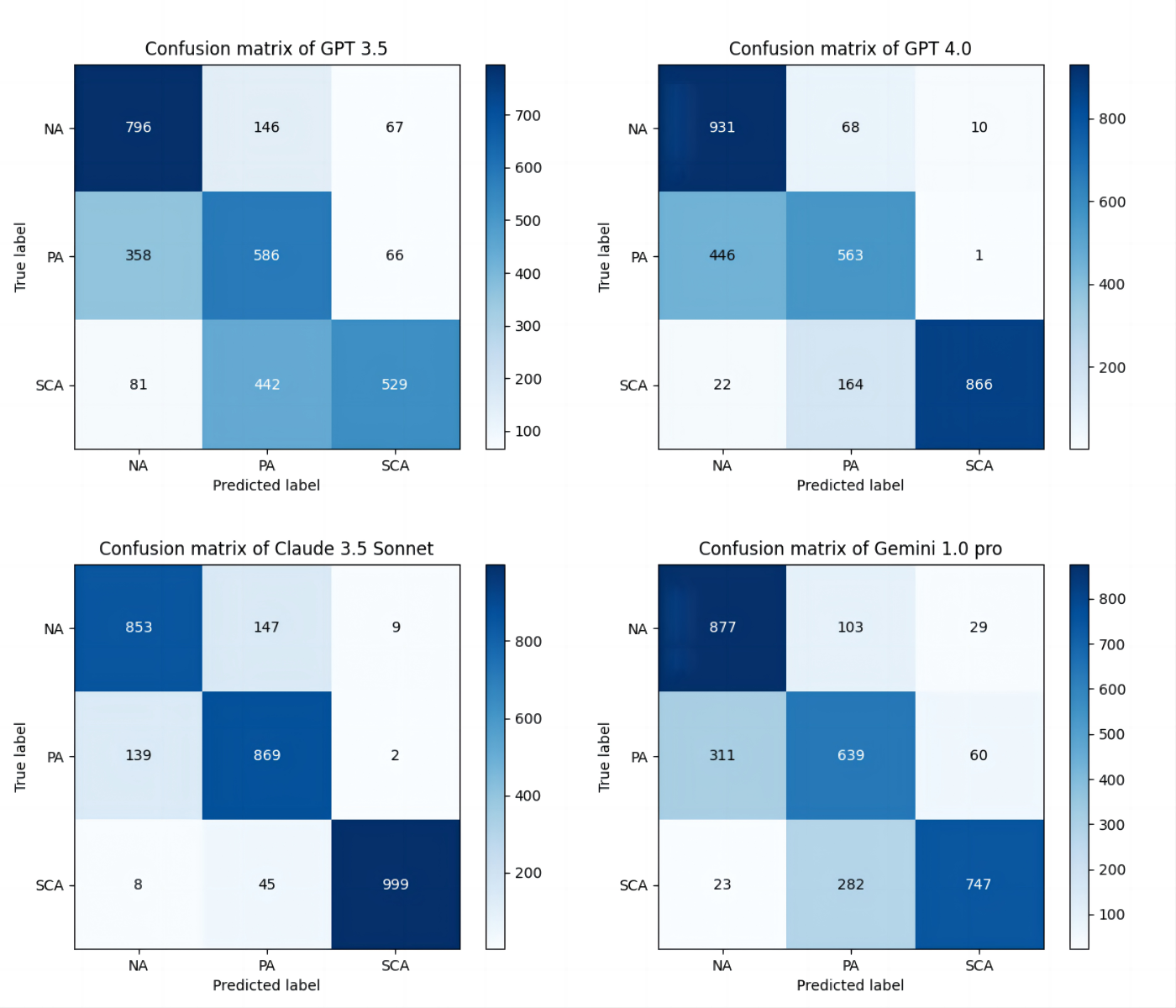}
\caption{Confusion matrices of the selected decoder-only models}
\label{confusion_matrix_gpt}
\end{figure}

The confusion matrix of ChatGPT (based on the GPT-3.5 model and the GPT-4 model), Claude (based on the Claude 3.5 Sonnet model), and Gemini (based on the Gemini 1.0 pro model) are shown in Figure \ref{confusion_matrix_gpt}. The accuracy, precision, recall, and f1-score of the selected decoder-only models are shown in Table \ref{Performance of the three decoder-only models}.
Specifically, if we consider whether the selected decoder-only models can be used to identify assumptions but not further distinguish between SCAs and PAs, that is, if we simply consider both SCAs and PAs as assumptions and therefore treat the problem as a binary classification problem (i.e., assumptions and non-assumptions), then the selected decoder-only models perform better compared to the multiclass classification problem (e.g., ChatGPT (based on the GPT-3.5 model) gets an f1-score of 0.7711, compared to 0.6211).

\textbf{Summary}: Compared to the results of RQ1 and RQ2, the worst performance of the selected decoder-only models is better than the performance of the trained seven non-transformers based models, while the best performance of the selected decoder-only models is worse than the performance of the trained ALBERT model.

\begin{table}[!htbp]
\scriptsize
\centering
\caption{Performance of the selected decoder-only models of assumption identification on the AssuEval dataset}
\label{Performance of the three decoder-only models}
\begin{tabular}{cccccc}
\toprule
Model & Version & Accuracy & Precision & Recall & F1-score \\
\midrule
\multirow{2}{*}{ChatGPT} & GPT-3.5 & 0.6223 & 0.6476 & 0.6240 & 0.6211 \\
& GPT-4 & 0.7685 & 0.7870 & 0.7678 & 0.7650 \\
Claude & 3.5 Sonnet & \cellcolor{lightgray}0.8860 & \cellcolor{lightgray}0.8870 & \cellcolor{lightgray}0.8851 & \cellcolor{lightgray}0.8858 \\
Gemini & 1.0 pro & 0.7369 & 0.7473 & 0.7373 & 0.7366 \\
\bottomrule
\end{tabular}
\end{table}

\section{Discussion}\label{discussion}
\subsection{Performance of the models}
The selected models have different strategies to identify assumptions. Since transformers based models have been shown to be more efficient than non-transformers based models in many classification tasks, it is not surprising that the ALBERT model performed better than the seven non-transformers based models in assumption identification. 
However, the performance of the selected decoder-only models in this study is worse than the performance of the ALBERT model. ChatGPT based on the GPT 3.5 model has only an f1-score of 0.6211, which is much worse than the f1-score (0.9584) achieved by the ALBERT model. One reason is that the decoder-only models we used in this study were trained by the companies for general purposes, while the ALBERT model was trained based on the AssuEval dataset. Fine-tuning the decoder-only models specifically for assumption identification could improve their performance in this task.
Furthermore, although the ALBERT model achieved the highest performance, it has limitations, such as the length of the input. In this study, if the length of a sentence is more than 128 words, it will be truncated when using the ALBERT model to identify assumptions. Therefore, the ALBERT model trained in this study can only be used to identify assumptions at the sentence level. Compared to the ALBERT model, the length limit of the selected decoder-only models is much higher (e.g., 4096 tokens, including both input and output of ChatGPT based on the GPT-3.5 model). In the context of decoder-only models, a token is a chunk of text that can be as short as one character or as long as one word, depending on the language and context. Tokens include words, punctuation marks, and other elements. For example, if the input sentence is ``How are you?'', then the models count it as [``How'', ``are'', ``you'', ``?''], which has four tokens. Therefore, the models can be used to identify assumptions at the sentence level, paragraph level, or even document level.

\subsection{Reasons of low performance of the decoder-only models}
The non-transformers based models and the ALBERT model are more like a black box, and we do not know the rationale of how they identified the assumptions. 
Differentially, because of the chatting nature of the selected decoder-only models, we can ask them to provide not only the types of each sentence, but also further analysis behind the answers, such as why ChatGPT identified a sentence as a specific type. It is important to remember that the selected decoder-only models are actually language models and that they cannot ``think'' or present cognitive abilities in the human sense \cite{Shanahan2023}. Such models operate by identifying patterns in the data and making predictions based on the statistical properties of the text on which they have been trained. When we refer to the models that provide a ``rationale'', it is a simplified way to describe the model's process of generating text based on the input it receives and the vast amount of data on which it was trained. This does not equate to the model understanding or reasoning in the way humans do \cite{Shanahan2023}.

After analyzing the data, we can see that the selected decoder-only models have made many errors in the classification task. We list ten examples of incorrect classification with the rationale provided by ChatGPT (based on the GPT-3.5 model) in Table \ref{Rational of assumption identification by ChatGPT}.
For E1, E2, E5, E7, and E9, the raw data qualifies the SCA\_IC1 criterion, which should be classified as SCA. In E1, although ChatGPT identified the keyword ``assuming'', it misunderstood the sentence as a question. In E2, ChatGPT considered the sentence as ``addressing the reader'', which is not true. In E5, the sentence includes ``assume'', which is an assumption-related keyword, while ChatGPT said that the sentence does not use assumption-related keywords. In E7, ChatGPT explained the assumption as speculation and considered that it has to be confirmed. In E9, ChatGPT only saw part of the sentence: ``something causes problems''. However, the sentence explicitly claimed an assumption that TensorFlow Probability's JAX backend assumes properties like `numpy.int32 == tf.int32' hold.
For E3, ChatGPT identified an assumption-related keyword: ``assumes'' in the sentence, which does not exist.
For E4 and E6, the sentence qualified the PA\_IC1 criterion, while ChatGPT did not identify anything in the sentence.
For E8 and E10, they are statements of facts, while ChatGPT considered them as assumptions. The reason is that sometimes ChatGPT would imagine the context of the sentence, even though we clearly asked it not to do that.

{
\scriptsize
\begin{longtable}[!htbp]{p{0.05\columnwidth}p{0.34\columnwidth}p{0.06\columnwidth}p{0.07\columnwidth}p{0.35\columnwidth}}
\caption{Rational of assumption identification by ChatGPT}\label{Rational of assumption identification by ChatGPT}\\  \hline
\toprule
\textbf{No.} & \textbf{Raw data} & \textbf{Ground truth} & \textbf{ChatGPT} & \textbf{Rationale by ChatGPT}\\
\midrule
{E1} & {I am assuming there's not a memory problem since on cpu it runs fine and starts training rather quickly, so I can see the progress plus when running on cpu the allocated RAM doesn't go beyond 3GB.} & {SCA} & {PA} & {The sentence includes the term ``assuming,'' but it involves a question about the existence of a memory problem. Since the question is not explicitly an assumption, it falls into the category of Potential Assumption (PA).}\\
{E2} & {@petewarden Assume you're new to TensorFlow.} & {SCA} & {NA} & {The sentence contains the term ``Assume,'' but in this context, it appears to be addressing the reader rather than making an assumption. Therefore, it is classified as Not an Assumption (NA).}\\
{E3} & {Remove ``at least 2D'' rank expansion in fit/predict/evaluate.} & {NA} & {SCA} & {The sentence mentions ``assumes'' and describes a specific assumption about the rank expansion. It meets the inclusion criteria for SCAs.}\\
{E4} & {...I think it would be just silly to merge the base class `RecurrentAttentionCellWrapperABC' with the specific implementation `MixtureOfGaussian1DAttention' as it describes and implements a highly relevant and general abstraction level for attention mechanisms.} & {PA} & {NA} & {The sentence discusses the merging of code without explicitly containing assumptions or potential assumptions.}\\
{E5} & {I assume it means the code that Theano is generating has if statements nested too deeply.} & {SCA} & {PA} & {The sentence indicates the author's assumption or interpretation (``I assume it means the code that Theano is generating has if statements nested too deeply''). It does not use assumption-related keywords but expresses a potential assumption.}\\
{E6} & {Now it seems to be a pain point to repeat these image\_dim\_ordering() tests.} & {PA} & {NA} & {The sentence discusses a pain point related to testing but does not convey any assumptions or expectations. It is classified as Not an Assumption (NA).}\\
{E7} & {This sort of question is better asked on stackoverflow, assuming that it's user error and not a completely broken model.} & {SCA} & {PA} & {The sentence involves a question and speculation (``assuming that it's user error''), indicating a potential assumption that needs further confirmation.}\\
{E8} & {I have tried to keep the output format for the new function as close to the existing `categorical\_accuracy()' function as possible.} & {NA} & {SCA} & {The sentence explicitly mentions keeping the output format close to an existing function, indicating an assumption about maintaining similarity with the categorical\_accuracy() function.}\\
{E9} & {This largely works fine, but causes problems for TensorFlow Probability's JAX backend, which assumes properties like `numpy.int32 == tf.int32' hold.} & {SCA} & {PA} & {The sentence describes a situation (``causes problems'') without explicitly stating assumptions. It suggests a potential issue without confirming assumptions.}\\
{E10} & {- generalized for any lahead/tsteps combination} & {NA} & {PA} & {The sentence is a comment related to generalization and does not explicitly state assumptions. It indicates a consideration without confirming assumptions.}\\
\bottomrule
\end{longtable}
}

Considering the Claude Sonnet 3.5 model, though it achieved the best performance in the selected decoder-only models, it also made many mistakes in identifying assumptions. We list ten examples of incorrect classification with the rationale provided by the Claude Sonnet 3.5 model in Table \ref{Rational of assumption identification by Claude}.
For E1, E5, E7, and E10, the raw data qualifies the inclusion criteria for SCAs, which should be classified as SCA. In E1 and E5, Claude did not identify the keyword ``assume''. In E7, the original sentence indicates that there exist traditional compiler ``fastmath'' assumptions, whereas Claude did not identify it. In E10, though the original sentence is a question, the question is about the validity of the assumption instead of whether something is an assumption or whether to make an assumption, i.e., the assumption exists, which qualifies SCA\_IC3. Although Claude identified the question and also indicated that the question is about the validity of an existing assumption, it seems that Claude twisted the meaning of SCA\_EC4 and confused SCA\_IC3 with SCA\_EC4.
For E2 and E4, Claude said that there was an assumption-related keyword in the sentence that does not exist.
For E3, the sentence is simply a question regarding whether something is an assumption, while Claude misused SCA\_IC3.
For E6, the sentence is a statement of facts, while Claude considered it as a PA. The reason is that sometimes Claude would imagine the context of the sentence, even though we clearly asked it not to do that.
For E8 and E9, the sentences qualified the PA\_IC1 criterion, while Claude did not identify anything in the sentence.

Although the performance of Claude is much better than that of ChatGPT in identifying assumptions, they made similar types of mistakes. For example, they sometimes make things up and build up their description based on the things made up. This finding is consistent with many other studies (e.g., \cite{King2023}\cite{Lingard2023}\cite{Hacker2023}).

{
\scriptsize
\begin{longtable}[!htbp]{p{0.05\columnwidth}p{0.34\columnwidth}p{0.06\columnwidth}p{0.07\columnwidth}p{0.35\columnwidth}}
\caption{Rational of assumption identification by Claude}\label{Rational of assumption identification by Claude}\\  \hline
\toprule
\textbf{No.} & \textbf{Raw data} & \textbf{Ground truth} & \textbf{Claude} & \textbf{Rationale by Claude}\\
\midrule
{E1} & {The PR has been updated, I assume it will fix the error.} & {SCA} & {PA} & {This sentence contains a guess about a future event (fixing an error). It meets PA inclusion criterion 1, as it's not an SCA and contains a valid guess that is not a question, condition, or alternative.}\\
{E2} & {And i suppose that it doesn't work in sequential models since they are not named.} & {PA} & {SCA} & {This sentence contains an explicit assumption using the word ``suppose''. It meets the inclusion criteria 1 for SCAs, as it contains a valid assumption using an assumption-related keyword.}\\
{E3} & {Do you assume that the entire dataset to be shuffled fits in memory?} & {NA} & {SCA} & {This sentence uses the keyword ``assume'' in a question about assumptions. It meets the inclusion criteria 3 for SCAs, as it contains an assumption in a question, and the question is not about whether something is an assumption or whether to make an assumption.}\\
{E4} & {I understand that this warning is to potentially spot channels\_first/channels\_last mixups, but the false positives [are causing problems](https://github.com/keras-team/keras/issues/11538).} & {PA} & {SCA} & {This sentence explicitly uses the word ``assumption'' to describe an expected behavior. It meets the inclusion criteria 1 for SCAs, containing a valid assumption using an assumption-related keyword.}\\
{E5} & {You could assume the fully-connected layers have as many multiplications as weights (plus bias), and that the pooling layers have none.} & {SCA} & {PA} & {This sentence contains a PA, suggesting a possibility about how to count multiplications. It meets the inclusion criteria 1 for PAs, as it contains a valid suggestion that is not a question, condition, or alternative.}\\
{E6} & {I am having second thoughts on it.} & {NA} & {PA} & {This sentence expresses uncertainty or reconsideration, which can be interpreted as a potential assumption about the speaker's changing opinion. It meets the inclusion criteria for PAs.}\\
{E7} & {What we want for an ML compiler is different from traditional compiler ``fastmath'' assumptions.} & {SCA} & {PA} & {This sentence meets PA inclusion criterion 1. It expresses an opinion or expectation about what is wanted for an ML compiler, which could be considered a potential assumption.}\\
{E8} & {ValueError: The model expects 0 input arrays, but only received one array. Found: array with shape (32, 299, 299, 3)} & {PA} & {NA} & {This sentence does not contain any SCAs or PAs. It's simply stating an error message without any assumptions or related elements.}\\
{E9} & {Maybe an example would be helpful, too.} & {PA} & {NA} & {This sentence does not contain any SCAs or PAs. It's a suggestion without expressing an assumption or potential assumption.}\\
{E10} & {Is my assumption wrong on the time complexity?} & {SCA} & {NA} & {While this sentence contains the word ``assumption'', it's asking a question about whether an assumption is wrong. This meets the exclusion criteria 4 for SCAs, as it's only asking about the validity of an assumption. It also doesn't qualify as a PA.}\\
\bottomrule
\end{longtable}
}

\subsection{Implications to researchers and practitioners}
\textbf{Assumption documentation}. In the data collection phase of this study, we can see that the assumptions are scattered in various sources. Therefore, we recommend that researchers investigate and design best practices or frameworks to document assumptions in DL framework projects. For example, if assumptions and their impacts can be explicitly stated with accepted effort, it would make it easier for practitioners to understand and manage these assumptions over time.
From the perspective of practitioners, they need to emphasize clear and thorough documentation of assumptions in, for example, code, comments, issues, and project discussions. This will help the development team and future contributors understand the rationale behind certain design decisions and anticipate potential risks caused by assumptions.

\textbf{Assumption classification}. As shown in the AssuEval dataset, there are various types of assumptions in DL framework development. In our earlier work, we identified 3,084 SCAs from code comments in over 50,000 files of nine popular DL frameworks (e.g., TensorFlow, Keras, and PyTorch) on GitHub \cite{Yangsca2021}, and analyzed the types of the identified SCAs. Researchers could work on developing a comprehensive taxonomy for classifying assumptions in DL framework development based on the taxonomy in our previous work with the dataset in this study. 

\textbf{Assumption evaluation}. We also observed that there are critical assumptions that significantly impact the project and less crucial ones that might have minimal effects. Researchers need to develop approaches to assess the potential impact of different assumptions. This can help prioritize resources towards addressing the most critical assumptions that could influence decision making, code structure, and overall project success.
In addition, in the data collection phase of this study, we can see that new assumptions arise throughout the development life cycle. Since assumptions may impact other assumptions and other types of software artifacts, practitioners need to review existing assumptions to check, for example, if they are valid. Regular reviews ensure that the team remains proactive in managing assumptions. Finally, since not all assumptions are important, practitioners need to understand which assumptions have a substantial impact on project success and focus their resources on managing and validating those assumptions appropriately.

\textbf{Fine-tuning the pre-trained models}. In this study, ALBERT has shown the best performance in the identification of assumptions at the sentence level. Researchers can further investigate and develop fine-tuning strategies to improve the performance of models. Moreover, this finding might be generalized to decision identification, technical debt identification, etc. We need further exploration of transformer models in both assumption identification and other related classification tasks in DL framework development.

\textbf{Awareness of assumptions}. Practitioners should be actively aware of the assumptions underlying the DL framework projects. This awareness is crucial to make informed decisions, anticipate potential challenges, and ensure the robustness of the developed solutions. For example, when users raise issues of the system, practitioners are encouraged to discuss and identify if there are assumptions behind the issues, and analyze their impacts on existing assumptions or other types of software artifacts. The results of the assumption analysis can be used for more robust decision-making and solutions of the issues.

\textbf{Model selection}. The selection of models is important, which should be based on the specific needs of the task at hand. According to the study findings, researchers and practitioners can lean toward using the ALBERT model when dealing with sentence-level tasks, such as assumption identification. However, researchers and practitioners should also be aware that although the ALBERT model is a lightweight transformer-based model, it has certain computational demands. Researchers and practitioners should weigh the benefits of performance against the complexity and resource requirements of these models, especially in practical software development. In addition, researchers and practitioners can also consider incorporating the ALBERT model into NLP pipelines for tasks that involve analyzing or categorizing individual sentences.
Despite the relatively lower performance of the selected decoder-only models (with the best f1-score of 0.8858 of the selected decoder-only models compared to ALBERT's 0.9584) in classifying assumptions within our study, researchers and practitioners may still find it beneficial to incorporate the selected decoder-only models for understanding the context of assumptions (e.g., the rationale of making certain assumptions). Their capacity to generate responses and comprehend the context of assumptions can effectively be used in assumption extraction (e.g., extraction of the relationships between assumptions and requirements, design decisions, etc.) and assumption reasoning (e.g., reasoning of the conflicts between assumptions). This is the next step in the analysis of sentence-level assumptions, which can complement the strengths of the ALBERT model, allowing for a more comprehensive understanding of assumptions.

\section{Threats to Validity}\label{threats}
We discuss the potential threats to the construct validity, external validity, and reliability of this study according to the guidelines proposed by Runeson and Höst \cite{Runeson2012}. 
This work does not study causality, and internal validity is not considered.

\textbf{Construct validity} reflects the degree to which the research questions and the operational measures studied are consistent \cite{Runeson2012}.
A potential threat comes from the construction of the AssuEval dataset~\cite{yangpackage2023}. For example, researchers might have a different understanding of the assumption concept, leading to different labels for the same sentence. To alleviate this threat, we conducted a pilot labeling on the TensorFlow and Keras repositories, and the criteria were iteratively refined during the pilot study. In the formal labeling process, all authors reviewed the results. To eliminate personal bias, conflicts and disagreements about labeling results were discussed and resolved among the authors.
Moreover, decoder-only models are context-dependent. For the same question, the answers may be different in various contexts. Therefore, there is a risk that the selected decoder-only models may forget the definitions and criteria when identifying assumptions in sentences sequentially. To reduce this threat, the sentences provided to the selected decoder-only models were separated in different chats with the same context.
Finally, as clarified in Section \ref{criteria}, we did not consider a sentence with a question mark as a question if it was not in the standard form. Contributors to the selected projects may not have English as their first language, so they might not write in the standard way. In the current design, there may be this kind of data that is mislabeled. However, in OSS projects, it is difficult to communicate with the contributors to confirm their assumptions, and therefore we have to develop explicit rules for labeling assumptions. Otherwise, the labeling process would be subjective, which may cause serious threats to the study design (e.g., cannot replicate the study by other researchers). We further checked the AssuEval dataset, and 1,015 sentences have at least one question mark, in which 396 sentences are not in the standard form (out of 15,354, 2.58\%), which is a small part.

\textbf{External validity} concerns the generalization of the findings \cite{Runeson2012}. 
In this study, we only selected two popular DL framework projects to evaluate the performance of the selected models in identifying assumptions. However, the two projects, TensorFlow and Keras, are representative in DL frameworks, and therefore the results can be partially generalized to other DL framework projects or those DL applications that are based on these two DL frameworks. The results may not be applicable to other types of software projects (e.g., non-ML projects). Replicating the study on other types of software projects can provide more evidence and help to generalize the results of this study, which we consider as future work.
Moreover, assumptions may be scattered in different sources (e.g., commits, PRs, issues, code comments, and documentation), while in this study, we selected three of them: commits, PRs, and issues). To reduce the threat, we followed the guidelines proposed by Stiennon \textit{et al.} \cite{Stiennon2022} to calculate and construct a representative dataset used to evaluate the performance of the selected models in the identification of assumptions. The results on the dataset can be generalized to the entire TensorFlow project.
Finally, the selected models are representative, and the results can be generalized to other similar models.

\textbf{Reliability} focuses on whether the study would produce the same results when other researchers replicate it \cite{Runeson2012}.
To alleviate this threat, we iteratively discussed the research protocol that was confirmed by all authors, and we provide the details of the study design in Section \ref{design}. To facilitate replication of our work and mitigate threats, we provide a replication package, which is available online \cite{yangpackage2023}.
Moreover, the selected decoder-only models were built on DL neural networks. How they generate answers for a specific question is a black-box for their users. Therefore, we repeated our questions for ChatGPT (based on the GPT-3.5 model) and found that for the same question in the same context, although ChatGPT did not provide the exact same answer every time, the meaning of the answers was similar. This improves the reliability of the study. However, we note that the selected decoder-only models have evolved over time and that users may not be able to access their historical versions. Therefore, for future versions of the selected decoder-only models, they may have different answers to the questions, which can introduce threats to reliability.

\section{Conclusions} \label{conclusions}
Stakeholders constantly make assumptions about the development of DL frameworks. These assumptions are related to various types of software artifacts (e.g., requirements, design decisions, and technical debt) and can turn out to be invalid, leading to system failures. Existing approaches and tools for the management of assumptions usually depend on the manual identification of assumptions. However, assumptions are scattered in various sources (e.g., code comments, commits, pull requests, and issues) of DL framework development, and manually identifying assumptions has high costs (e.g., time and resources). 

To overcome the issues of manually identifying assumptions in DL framework development, we first constructed the AssuEval dataset, and then explored the performance of seven non-transformers based models (e.g., Support Vector Machine, Classification and Regression Trees), the ALBERT model, and three decoder-only models (i.e., ChatGPT, Claude, and Gemini) for identifying assumptions on the AssuEval dataset~\cite{yangpackage2023}. 
The study results show that ALBERT achieves the best performance (f1-score: 0.9584) for identifying assumptions on the AssuEval dataset, which is much better than the other models (the 2nd best f1-score is 0.8858, achieved by the Claude 3.5 Sonnet model). Though ChatGPT, Claude, and Gemini are popular models, we do not recommend using them to identify assumptions in DL framework development because of their low performance. Fine-tuning ChatGPT, Claude, Gemini, or other language models (e.g., Llama3, Falcon, and BLOOM) specifically for assumptions might improve their performance for assumption identification.

For the next steps, we consider the following directions:
(1) Types of assumptions: Implement the trained ALBERT model in the Assumption Miner tool and use it to identify and classify SCAs and PAs in DL framework projects. The classification can be based on various dimensions, such as the content and validity of assumptions, whether an assumption is violated, or whether an assumption has details. This helps us to understand the types of assumptions that exist in projects, which can be the basis for a further detailed analysis of these assumptions.
(2) Context analysis: In this study, we only considered assumptions at the sentence level since the focus is only on identifying assumptions without further analysis. However, the context of assumptions can provide additional information that can help stakeholders better understand, for example, why stakeholders made such assumptions, which is useful to explore the rationale for making assumptions in the development of DL frameworks.
(3) Characteristic analysis: Analyze the characteristics of the assumptions in DL framework projects. For example, develop methods and tools to analyze the relationships between assumptions and other types of related software artifacts. This helps stakeholders understand the importance of assumptions in projects and how assumptions impact each other and other types of software artifacts. Another example is to investigate the addition, update, transformation, and deletion of assumptions. This helps stakeholders to be aware of how their assumptions evolve over time in projects (e.g., from valid to invalid).
(4) Replicating the study to include more projects to further generalize the results of this study.

\section*{Acknowledgments} 
This work is funded by Shenzhen Polytechnic University with Grant No. 6022312043K, State Key Laboratory for Novel Software Technology at Nanjing University with Grant No. KFKT2022B37, the National Natural Science Foundation of China (NSFC) with Grant No. 62172311, and the Special Fund of Hubei Luojia Laboratory.

\bibliography{mybibfile}
\end{document}